\documentclass[12pt]{article}
\usepackage{amsmath}
\usepackage{amsthm}
\usepackage{amssymb}
\usepackage{mathtools}
\usepackage{graphicx,psfrag,epsf}
\usepackage{enumerate}
\usepackage{natbib}
\usepackage{url} 
\usepackage{booktabs}
\usepackage{color}
\usepackage{array}
\usepackage{multirow}
\usepackage{algorithm}
\usepackage{algpseudocode}

\begin{document}
	
\def\spacingset#1{\renewcommand{\baselinestretch}%
	{#1}\small\normalsize} \spacingset{1}
	
\title{\bf Learning Signal Subgraphs from Longitudinal Brain Networks with Symmetric Bilinear Logistic Regression}

\author{Lu Wang \\
	Department of Statistics, Central South University \\
	and \\
	Zhengwu Zhang \\
	Department of Biostatistics and Computational Biology, University of Rochester
}
\maketitle

\bigskip
\begin{abstract}
	Modern neuroimaging technologies, combined with state-of-the-art data processing pipelines, have made it possible to collect longitudinal observations of an individual's brain connectome at different ages. It is of substantial scientific interest to study how brain connectivity varies over time in relation to human cognitive traits. In brain connectomics, the structural brain network for an individual corresponds to a set of interconnections among brain regions. We propose a symmetric bilinear logistic regression to learn a set of small subgraphs relevant to a binary outcome from longitudinal brain networks as well as estimating the time effects of the subgraphs. We enforce the extracted signal subgraphs to have clique structure which has appealing interpretations as they can be related to neurological circuits. The time effect of each signal subgraph reflects how its predictive effect on the outcome varies over time, which may improve our understanding of  interactions between the aging of brain structure and neurological disorders. Application of this method on longitudinal brain connectomics and cognitive capacity data shows interesting discovery of relevant interconnections among a small set of brain regions in frontal and temporal lobes with better predictive performance than competitors.
	
\end{abstract}
\noindent%
\vfill

\newpage
\spacingset{1.45} 

\section{Introduction}
\label{sec:intro}

In this article, we study the methods for predicting a binary outcome $y_i$ from a sequence of longitudinal network-valued variables $\{W_{i}^{(s)}:s=1,\dots,T_{i}\}$ for $n$ subjects, where each observed network is undirected without self loops and hence each $W_i^{(s)}$ is a $V\times V$ symmetric matrix with zero diagonal. The motivation of this study is from the increasing interest of understanding brain connectomics and its relation to brain functions. With advanced neuroimaging technologies, more and more large neuroscience studies start collecting longitudinal brain scans, e.g., the Alzheimer's Disease Neuroimaging Initiative (ADNI) \citep{jack2008alzheimer} and the recent Adolescent Brain Cognitive Development project \citep{casey2018adolescent}. We extracted a subset of subjects in ADNI according to Lin et al.'s study \citep{lin2017cingulate} of successful cognitive aging, consisting of longitudinal structural brain networks over a 5-year span for 40 supernormals and 45 cognitively normal controls with matched age, gender and education. 
Supernormals refer to the older adults who do not exhibit the expected reduction in cognition, but have superior cognitive performance during aging \citep{lin2017cingulate}. 

\begin{figure}[ht]
	\centering
	\begin{tabular}{cc}
	\hspace{-10pt}\includegraphics[width=.5\textwidth]{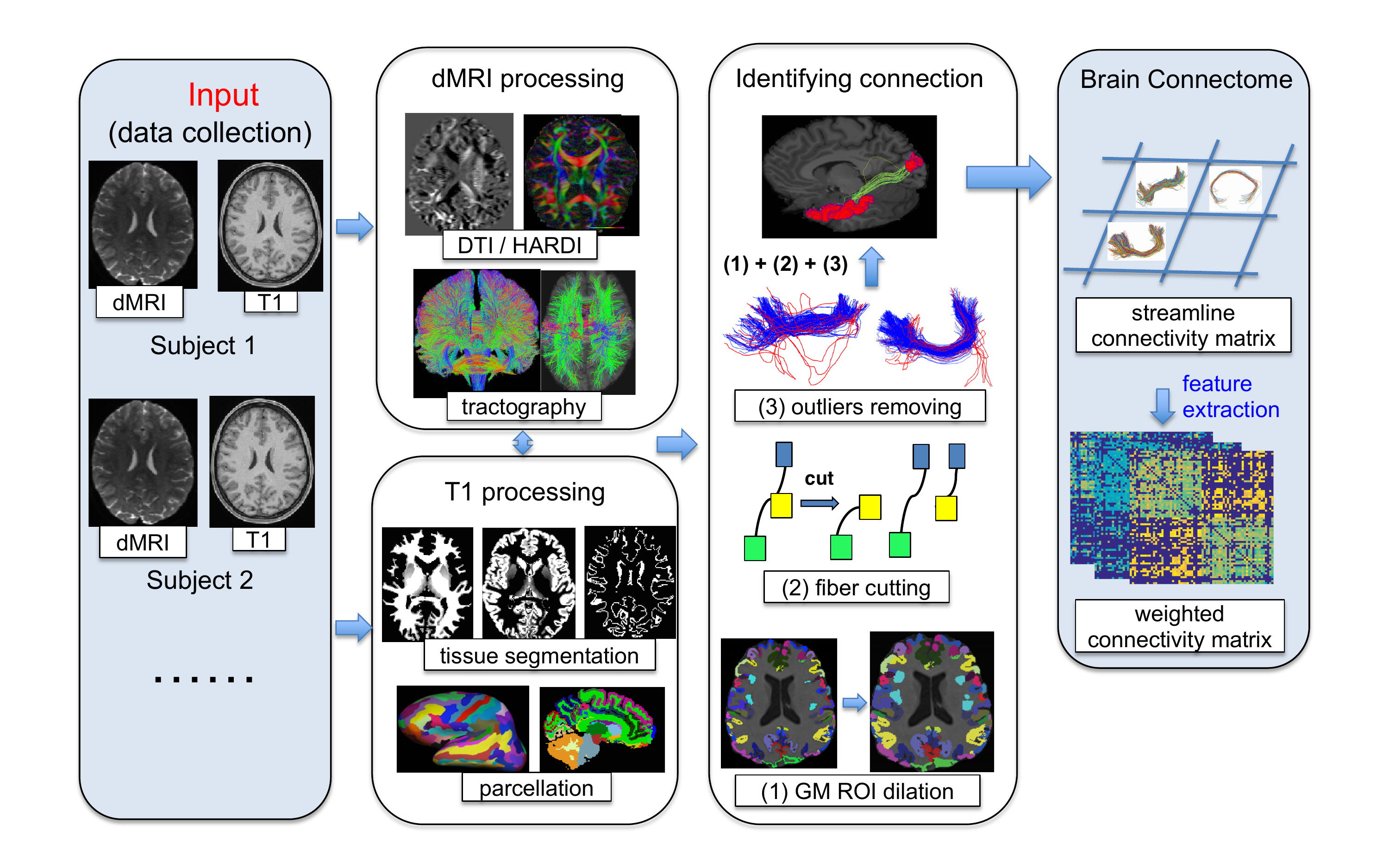} &
	\includegraphics[width=.5\textwidth]{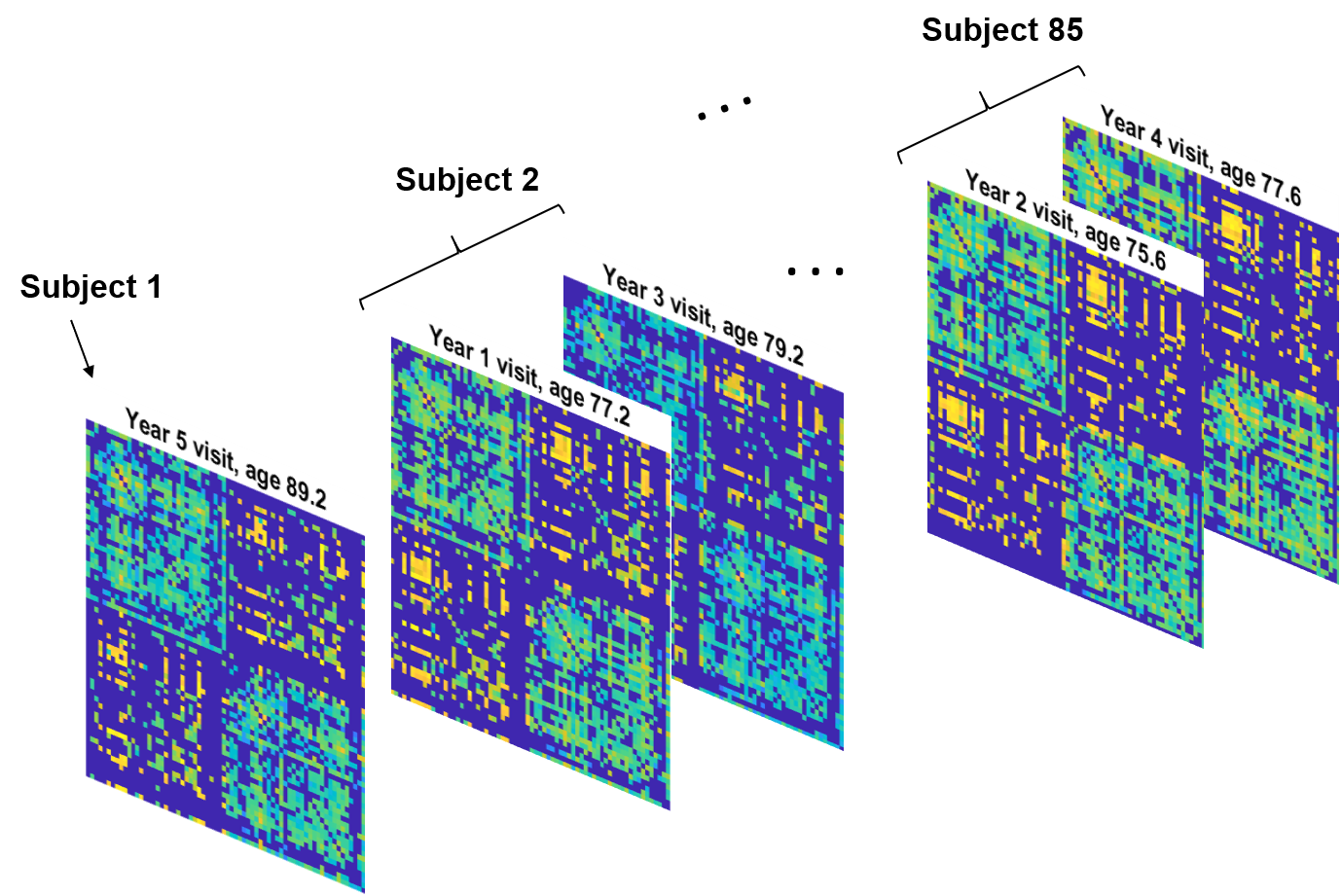} \\
	(a) & (b) 
      \end{tabular}

	\caption{Panel (a) shows our structural brain network extraction pipeline and panel (b) shows the profile of longitudinal brain networks for each individual in the dataset.}
	\label{data-profile}
\end{figure}

Using state-of-the-art connectome extraction pipeline \citep{zhang2018mapping}, we can extract structural brain networks from diffusion MRI (dMRI) data. The structural brain network for an individual corresponds to a set of white matter connections among predefined brain regions. In our data,
the number of observed brain networks for each individual ranges from 1 to 5 since not all the participants visited every year during the 5-year program. Each brain network is characterized by a weighted adjacency matrix where each element denotes the connectivity strength of neural fibers between a pair of brain regions. Figure \ref{data-profile} left panel shows the preprocess of recovering structural connectomes from   dMRI data, and the right panel shows a profile of the dataset we used. 

Our scientific goal is to study the relationship between successful cognitive aging and structural brain networks in hope of finding neurologically interpretable structural markers in brain connectome predictive of supernormals. By neurologically interpretable markers we mean a set of small outcome-relevant subgraphs with certain structure in the brain network. For example the connections within each subgraph should share some common nodes (brain regions) or form certain anatomical circuits. With age information of the subjects, we also aim to estimate the age effect of each identified signal subgraph, that is the dynamic importance of each subgraph in preserving cognition, which may improve cognition diagnosis and help understand the role of aging in normal and diseased brains.

The signal subgraph learning is basically a variable selection problem where the number of predictors - corresponding to the number of connections in the longitudinal brain networks - could easily exceed the number of observations (subjects). One typical approach to this large $p$ small $n$ problem would be some generalized linear model with certain regularization, such as LASSO \citep{tibshirani1996regression}, Elastic-Net \citep{zou2005elastic-net} and SCAD \citep{fan2001variable}. But these methods do not guarantee any structure among the selected connections in the brain network, making the results hard to interpret neurologically. In addition, these approaches require first transforming each adjacency matrix into a long vector, which could induce ultra high dimensionality for big networks and ignore non-Eucliean structure of networks. 

The longitudinal networks can be easily formed into tensors and thus tensor regression models \citep{zhou2013tensor, zhou2014regularized} provide an effective tool for our problem. But the weighted adjacency matrices of structural brain networks are symmetric, so additional symmetry constraints are required on the coefficient vectors in tensor regression. \cite{wang2019symmetric} propose a symmetric bilinear regression for estimating a set of small clique signal subgraphs from a network predictor. But they focus on a continuous response and do not consider the case of longitudinal network observations. In addition, the censoring of longitudinal networks prevents the direct application of existing tensor regression models. 

We propose to use a symmetric bilinear logistic regression (SBLR) model with partial time-varying coefficients and elastic-net penalty to learn a set of small clique subgraphs and estimate their age effects. The clique signal subgraphs selected by SBLR have more appealing interpretations than the results of unstructured variable selection tools especially in neurological studies, as many complex cognitive processes are often the product of coordinated activities among several brain regions. The estimated age effects demonstrate the dynamic impact on the outcome of the selected signal subgraphs, which may improve cognition diagnosis and development of treatment strategies for neurological disorders. SBLR is also very flexible in dealing with longitudinal network data, where each subject can have different number of observations at different time points. To adapt to the brain network data, the model puts symmetry constraints on the coefficient vectors of a tensor regression, which makes the block relaxation algorithm \citep{zhou2013tensor} not applicable. We therefore develop a coordinate descent algorithm for model estimation based on the success of R package \texttt{glmnet} \citep{friedman2010regularization}, which involves solving a sequence of conditional convex optimizations. 

The rest of the paper is organized as follows. The SBLR model with elastic-net penalty is proposed in Section 2. Section 3 describes the coordinate descent algorithm for model estimation and discusses model selection issues. Section 4 presents simulation studies demonstrating the advantages of our method over competitors. We apply SBLR model on the longitudinal brain network data in Section 5. Section 6 concludes.

\section{Symmetric Bilinear Logistic Regression (SBLR)}
\label{sec:model}

The data can be summarized as $\left( y_i, \{(W_i^{(s)}, g_{is}): s=1,...,T_i\} \right)$ for subject $i$, $i=1...,n$.  The binary response $y_i = 1$ denotes that subject $i$ is a supernormal and $y_i = 0$ for a normal control.  $T_i$ represents the number of network observations for subject $i$. $W_i^{(s)}$ denotes the structural brain network measured for subject $i$ at the $s$-th visit ($s=1,\dots,T_i$),  which is a $V\times V$ symmetric matrix with zero diagonal entries. $g_{is}$ is the age of subject $i$ when the network $W_i^{(s)}$ is observed. Our goal is to learn a set of small clique subgraphs from the brain network that are relevant to the outcome and estimate their age effects. 

We borrow ideas from the symmetric bilinear model \citep{wang2019symmetric} to adapt to a binary response with longitudinal network predictors by adding a logit link and making some of the coefficients time-varying. This leads to the following symmetric bilinear logistic regression (SBLR):
\begin{equation}
\begin{aligned}
& y_{i}\mid\{(g_{is},W_{i}^{(s)}):s=1,\dots,T_{i}\}\overset{ind}{\sim}\mbox{Bernoulli}(p_{i}),\ i=1,\dots,n,\\
& \mbox{logit}(p_{i})=\alpha_{0}+\sum_{h=1}^{K}\dfrac{1}{T_{i}}\sum_{s=1}^{T_{i}}\lambda_{h}(g_{is})\boldsymbol{\beta}_{h}^{\top}W_{i}^{(s)}\boldsymbol{\beta}_{h},
\end{aligned}
\label{SBLR-model}
\end{equation}
where $\boldsymbol{\beta_h} \in \mathbb{R}^V$ and $\lambda_h(\cdot)$ is a function, $\lambda_h:[a,b] \mapsto \mathbb{R}$, mapping from an interval $[a,b]$ to a scalar. Model \eqref{SBLR-model} assumes that the binary outcome $y_i$ of each individual follows an independent Bernoulli distribution given her longitudinal network observations and the corresponding age $\{(W_i^{(s)}, g_{is}): s=1,...,T_i\}$.

The coefficients in model \eqref{SBLR-model} are assumed to have $K$ components, where each component $\boldsymbol{\beta}_h\boldsymbol{\beta}_h^T$ selects a signal subgraph. 
For ease of interpretation, the logit link of \eqref{SBLR-model} can be written in matrix dot product form:
\begin{equation}
\mbox{logit}(p_{i})=\alpha_{0}+\dfrac{1}{T_{i}}\sum_{s=1}^{T_{i}}\sum_{h=1}^{K}\left\langle \lambda_{h}(g_{is})\boldsymbol{\beta}_{h}\boldsymbol{\beta}_{h}^{\top},W_{i}^{(s)}\right\rangle
\label{SBLR-matrix-prod} 
\end{equation}
where $\left\langle B,W\right\rangle =$ trace($B^{\top}W$) = vec$(B)^{\top}$vec($W$). From \eqref{SBLR-matrix-prod}, we can see that the nonzero entries in each component matrix $\boldsymbol{\beta}_{h}\boldsymbol{\beta}_{h}^{\top}$ locate an outcome-relevant clique subgraph in the network predictor. We do not let $\boldsymbol{\beta}_{h}$ vary with age in  \eqref{SBLR-model} for two reasons. First, we assume the subgraphs and the related brain regions associated with $y$ are stable across time for healthy adults (without cognitive impairment). The connection strengths among these brain regions could change over time, and their dynamic contribution to the outcome is captured by $\lambda_h(g)$. Note that $\lambda_h(\cdot)$  is also necessary to avoid constraining the coefficient matrix in \eqref{SBLR-matrix-prod} to be positive semi-definite (p.s.d). Second, we fix each $\boldsymbol{\beta}_{h}$ over time for model simplicity. Otherwise a random function for each element of $\boldsymbol{\beta}_{h}$ ($h=1,\dots,K$) will lead to intractable model estimation and overfitting issues.

Although individuals have repeated measures of her structural brain network at different time points in the dataset, we focus on estimating the common signal subgraphs predictive of $y$ as well as the age effects in model \eqref{SBLR-model} instead of their individual-specific counterparts due to the limited number of observations over time for each individual. In terms of the number of time points, longitudinal structural neuroimaging data are relatively rare, with most studies having less than three time points \citep{king2018longitudinal}. In our motivating data, each subject has no more than $5$ brain scans and around 87\% of the subjects only have one or two network observations. Therefore there are insufficient data to reliably estimate the intra-individual signal subgraphs or individual trajectories of age effects. However, the longitudinal network observations $\{W_{i}^{(s)}:s=1,\dots,T_{i}\}$ for each subject, though temporally correlated, could still contribute to inference of the common signal subgraphs $\{\boldsymbol{\beta}_{h}\boldsymbol{\beta}_{h}^{\top}\}$ and the population-level age effects $\{\lambda_{h}(g)\}$. So we utilize all the observed longitudinal network information for each individual in model \eqref{SBLR-model}. And since not all the subjects have the same number of network observations in the data, we divide the bilinear part by $T_i$ in the logit of \eqref{SBLR-model}.

Suppose that $y_i$ is related to some subgraph corresponding to $\boldsymbol{\beta}_{h}\boldsymbol{\beta}_{h}^{\top}$, within which the connection strengths of older adults with normal aging tend to decrease with age, while those of supernormals tend to maintain unchanged as they grow older. Then higher \emph{weight} should be put on the term $\boldsymbol{\beta}_{h}^{\top}W_{i}^{(s)}\boldsymbol{\beta}_{h}$ observed at an older age in predicting $y$. In this case, $\lambda_h(g)$ would be expected to increase with age and thus we could let $\lambda_h(g) = \rho_h g+\alpha_h$.  $\lambda_h(g)$ could also be assumed as a higher order polynomial function to enhance flexibility.
In simulations and applications, we let $\lambda_h(g)$ be up to second order, i.e. $\lambda_h(g) = \gamma_h g^2 + \rho_h g+\alpha_h$, for interpretation consideration and model simplicity, since higher order terms of age add difficulty to interpretation and are prone to overfitting. In addition, we rarely observed nonzero coefficient estimates for the quadratic terms of age in real applications with elastic-net penalty as shown in Section 4. 

Plug in the quadratic function $\lambda_h(g) = \gamma_h g^2 + \rho_h g+\alpha_h$ into the logit link of \eqref{SBLR-model}. We have
\begin{equation}
\begin{aligned}
\mbox{logit}(p_{i}) & =\alpha_{0}+\sum_{h=1}^{K}\left[\left\langle \alpha_{h}\boldsymbol{\beta}_{h}\boldsymbol{\beta}_{h}^{\top},\dfrac{1}{T_{i}}\sum_{s=1}^{T_{i}}W_{i}^{(s)}\right\rangle +\left\langle \rho_{h}\boldsymbol{\beta}_{h}\boldsymbol{\beta}_{h}^{\top},\dfrac{1}{T_{i}}\sum_{s=1}^{T_{i}}g_{is}W_{i}^{(s)}\right\rangle \right.\\
& \qquad\qquad\qquad\left.+\left\langle \gamma_{h}\boldsymbol{\beta}_{h}\boldsymbol{\beta}_{h}^{\top},\dfrac{1}{T_{i}}\sum_{s=1}^{T_{i}}g_{is}^{2}W_{i}^{(s)}\right\rangle \right].
\end{aligned}
\label{SBLR-logit1}
\end{equation}
Equation \eqref{SBLR-logit1} implies that under the quadratic assumption for $\lambda_h(g)$, the true covariates for each subject $i$ in model \eqref{SBLR-model} are not the raw longitudinal network observations, but the average of her networks, the weighted average of the networks by her age and squared age separately.  
To ensure both the identifiability of the model and sparsity of coefficient matrices $\left\{ \left(\alpha_{h}\boldsymbol{\beta}_{h}\boldsymbol{\beta}_{h}^{\top},\rho_{h}\boldsymbol{\beta}_{h}\boldsymbol{\beta}_{h}^{\top},\gamma_{h}\boldsymbol{\beta}_{h}\boldsymbol{\beta}_{h}^{\top}\right):h=1,\dots,K\right\} $, we penalize the magnitude of the lower-triangular entries in these coefficient matrices with the following elastic-net penalty:
\begin{equation}
\delta\sum_{h=1}^{K}\sum_{u=1}^{V}\sum_{v<u}\left[\eta\left(|\alpha_{h}|+|\rho_{h}|+|\gamma_{h}|\right)\cdot|\beta_{hu}||\beta_{hv}|+(1-\eta)(\alpha_{h}^{2}+\rho_{h}^{2}+\gamma_{h}^{2})\beta_{hu}^{2}\beta_{hv}^{2}/2\right]
\label{elanet-pen}
\end{equation}
where the overall penalty factor $\delta>0$ and $\eta \in [0,1]$ controlling the fraction of $L_1$ regularization. The entrywise regularization \eqref{elanet-pen} ensures the identifiability of the coefficient matrices $\left\{ \left(\alpha_{h}\boldsymbol{\beta}_{h}\boldsymbol{\beta}_{h}^{\top},\rho_{h}\boldsymbol{\beta}_{h}\boldsymbol{\beta}_{h}^{\top},\gamma_{h}\boldsymbol{\beta}_{h}\boldsymbol{\beta}_{h}^{\top}\right):h=1,\dots,K\right\}$ in model \eqref{SBLR-logit1}, which cannot be achieved by simply penalizing the $L_1$ norms or $L_2$ norms of the parameters $\left\{ \left(\alpha_{h},\rho_{h},\gamma_{h},\boldsymbol{\beta}_{h}\right):h=1,\dots,K\right\}$. Refer to \citet{wang2019symmetric} for a detailed discussion.

\section{Estimation}
\label{estimation}

The parameters in SBLR model \eqref{SBLR-model} are estimated by minimizing the loss function below:
\begin{equation}
\begin{aligned}
\mbox{Loss function} & =-\dfrac{1}{n}\sum_{i=1}^{n}ll_{i}+\sum_{h=1}^{K}\sum_{u=1}^{V}\sum_{v<u}\delta\left[\eta\left(|\alpha_{h}|+|\rho_{h}|+|\gamma_{h}|\right)\cdot|\beta_{hu}||\beta_{hv}|\right.\\
 & \quad\qquad\left.+(1-\eta)(\alpha_{h}^{2}+\rho_{h}^{2}+\gamma_{h}^{2})\beta_{hu}^{2}\beta_{hv}^{2}/2\right]
\end{aligned}
\label{loss-func}
\end{equation}
where $ll_i$ is the log-likelihood of subject $i$. Plugging in the logit link of \eqref{SBLR-model}, we have
\begin{align}
ll_{i} & =y_{i}\log(p_{i})+(1-y_{i})\log(1-p_{i}) \nonumber \\
 & =y_{i}\left(\alpha_{0}+\sum_{h=1}^{K}\dfrac{1}{T_{i}}\sum_{s=1}^{T_{i}}\lambda_{h}(\tilde{g}_{is})\boldsymbol{\beta}_{h}^{\top}\tilde{W}_{i}^{(s)}\boldsymbol{\beta}_{h}\right)+\log(1-p_{i}) \label{log-likelihood}
\end{align}
where $\lambda_{h}(\tilde{g}_{is})=\gamma_{h}\widetilde{g_{is}^{2}}+\rho_{h}\tilde{g}_{is}+\alpha_{h}$, and $\{\widetilde{g_{is}^{2}}\}$, $\{\tilde{g}_{is}\}$ are standardized $\{g_{is}^{2}\}$ and $\{g_{is}\}$ respectively. $\{\tilde{W}_{i}^{(s)}\}$ in \eqref{log-likelihood} are obtained by normalizing the observations to have mean 0 and variance 1 for each connection in the brain network. In this way, entries in the matrix predictors $\dfrac{1}{T_{i}}\sum_{s=1}^{T_{i}}\tilde{W}_{i}^{(s)}$, $\dfrac{1}{T_{i}}\sum_{s=1}^{T_{i}}\tilde{g}_{is}\tilde{W}_{i}^{(s)}$, and 
$\dfrac{1}{T_{i}}\sum_{s=1}^{T_{i}}\widetilde{g_{is}^{2}}\tilde{W}_{i}^{(s)}$ are roughly of the same magnitude. It is easy to recover age effects in the original scale through $\lambda_{h}(g_{is})=\gamma_{h}^{(o)}g_{is}^{2}+\rho_{h}^{(o)}g_{is}+\alpha_{h}^{(o)}$ with
\begin{align*}
\gamma_{h}^{(o)} & =\dfrac{\gamma_{h}}{\sigma_{2}}\\
\rho_{h}^{(o)} & =\dfrac{\rho_{h}}{\sigma_{1}}\\
\alpha_{h}^{(o)} & =\alpha_{h}-\rho_{h}\dfrac{\mu_{1}}{\sigma_{1}}-\gamma_{h}\dfrac{\mu_{2}}{\sigma_{2}}
\end{align*}
where $\mu_1$, $\mu_2$ are the means of $\{g_{is}\}$ and $\{g_{is}^{2}\}$ respectively, and $\sigma_1$, $\sigma_2$ their standard deviations.

Although there is scaling indeterminacy between $\lambda_h(g)$ and $\boldsymbol{\beta}_{h}\boldsymbol{\beta}_{h}^{\top}$ within each component such that 
\[
\lambda_{h}(g_{is})\boldsymbol{\beta}_{h}\boldsymbol{\beta}_{h}^{\top}=c_{h}\lambda_{h}(g_{is})\cdot\dfrac{1}{c_{h}}\boldsymbol{\beta}_{h}\boldsymbol{\beta}_{h}^{\top}
\]
for any $c_h \neq 0$, this scaling of $\lambda_{h}(g)$ does not change the \emph{argmax} or \emph{argmin} if $\lambda_{h}(g)$ is a quadratic function, or the linear trend if $\lambda_{h}(g)$ is a linear or constant function. In practice, we always report the estimated age effect $\lambda_{h}(g)$ after scaling $\boldsymbol{\beta}_{h}\boldsymbol{\beta}_{h}^{\top}$ so that the off-diagonal element with the largest magnitude is 1 for each nonempty component. This rule explicitly displays the age effect of the connection with the strongest predictive effect on the response in each signal subgraph.

 From \eqref{SBLR-model} we know that logit($p_i$) is a quadratic function of each $\boldsymbol{\beta}_h$, and 
\begin{align*}
\dfrac{\partial^{2}ll_{i}}{\partial\boldsymbol{\beta}_{h}\partial\boldsymbol{\beta}_{h}^{\top}} & =(y_{i}-p_{i})\cdot\dfrac{2}{T_{i}}\sum_{s=1}^{T_{i}}\lambda_{h}(g_{is})W_{i}^{(s)}\\
 & \qquad-p_{i}(1-p_{i})\cdot\dfrac{4}{T_{i}^{2}}\sum_{s=1}^{T_{i}}\sum_{t=1}^{T_{i}}\lambda_{h}(g_{is})\lambda(g_{it})W_{i}^{(s)}\boldsymbol{\beta}_{h}\boldsymbol{\beta}_{h}^{\top}W_{i}^{(t)}
\end{align*}
which may not be negative semi-definite. Therefore $ll_i$ is not a concave function of $\boldsymbol{\beta}_h$ when fixing the other parameters and the block relaxation algorithm for tensor regression \citep{zhou2013tensor} is not applicable. Since the networks in this case are undirected without self loops, the diagonal of each adjacency matrix $W_i^{(s)}$ are set to zero. Then the loss function \eqref{loss-func} is actually a convex function of each entry $\beta_{hu}$ in $\boldsymbol{\beta}_h$ when fixing the others, which makes the coordinate descent algorithm very appealing. The challenge then lies in deriving the analytic form update for each parameter due to the nonsmoothness of \eqref{loss-func}. In the following, we discuss the technical details of coordinate descent algorithm for model estimation.

\subsection{Updates for Entries in $\left\{ \boldsymbol{\beta}_{h}\right\} _{h=1}^{K}$}
\label{sec:update-beta}

Minimizing the loss function \eqref{loss-func} with respect to $\beta_{hu}$, the $u$-th entry of $\boldsymbol{\beta}_h$, given all the other parameters becomes:
\begin{equation}
\min_{\beta_{hu}}\ L(\beta_{hu})=-\dfrac{1}{n}\sum_{i=1}^{n}ll_{i}(\beta_{hu})+d_{hu}\left|\beta_{hu}\right|+e_{hu}\beta_{hu}^{2}/2
\label{optimize-beta-hu}
\end{equation}
where 
\begin{align}
d_{hu} & =\delta\eta\left(|\alpha_{h}|+|\rho_{h}|+|\gamma_{h}|\right)\sum_{v\neq u}|\beta_{hv}| \label{d-formula}\\
e_{hu} & =\delta(1-\eta)(\alpha_{h}^{2}+\rho_{h}^{2}+\gamma_{h}^{2})\sum_{v\neq u}\beta_{hv}^{2}. \label{e-formula}
\end{align}
The formulae \eqref{d-formula} - \eqref{e-formula} imply that the penalty factors $d_{hu}$ and $e_{hu}$ in \eqref{optimize-beta-hu} for $|\beta_{hu}|$ are related to the nonzero entries in $\boldsymbol{\beta}_h$ excluding $\beta_{hu}$. Hence $\beta_{hu}$ is more likely to be shrunk to zero if the current number of nonzero entries in $\boldsymbol{\beta}_h$ is large. This adaptive penalty will lead to a set of sparse vectors $\left\{ \boldsymbol{\beta}_{h}\right\} _{h=1}^{K}$ and hence a set of small signal subgraphs.

Note that
\begin{align}
\dfrac{\partial ll_{i}}{\partial\beta_{hu}} & =(y_{i}-p_{i})\cdot \dfrac{2}{T_{i}}\sum_{s=1}^{T_{i}}\lambda_{h}(\tilde{g}_{is})\tilde{W}_{i[u\cdot]}^{(s)}\boldsymbol{\beta}_{h} \label{1st-deri-beta}\\
\dfrac{\partial^{2}ll_{i}}{\partial\beta_{hu}^{2}} & =-p_{i}(1-p_{i})\left[\dfrac{2}{T_{i}}\sum_{s=1}^{T_{i}}\lambda_{h}(\tilde{g}_{is})\tilde{W}_{i[u\cdot]}^{(s)}\boldsymbol{\beta}_{h}\right]^{2} \leq 0. \label{2nd-deri-beta}
\end{align}
Therefore each log-likelihood $ll_i$ is a concave function of $\beta_{hu}$ when fixing the other parameters and hence the objective function $L(\beta_{hu})$ in \eqref{optimize-beta-hu} is a convex function of $\beta_{hu}$. 

Suppose the current estimate for $\beta_{hu}$ at iteration $t$ is $\beta_{hu}^{(t)}$ and also fix the other parameters at their current estimates. We will approximate  $-\dfrac{1}{n}\sum_{i=1}^{n}ll_{i}(\beta_{hu})$ in \eqref{optimize-beta-hu} by taking a second-order Taylor expansion at $\beta_{hu}^{(t)}$:
\begin{equation}
-\dfrac{1}{n}\sum_{i=1}^{n}ll_{i}(\beta_{hu})\approx L_{1}^{(t)}(\beta_{hu}) \triangleq c_{hu}^{(t)}+b_{hu}^{(t)}\left(\beta_{hu}-\beta_{hu}^{(t)}\right)+\dfrac{a_{hu}^{(t)}}{2}\left(\beta_{hu}-\beta_{hu}^{(t)}\right)^{2}
\label{2nd-approx-LL}
\end{equation}
where
\begin{align}
c_{hu}^{(t)} & =-\dfrac{1}{n}\sum_{i=1}^{n}ll_{i}(\beta_{hu}^{(t)}) \label{c-formula}\\
b_{hu}^{(t)} & =-\dfrac{1}{n}\sum_{i=1}^{n}\dfrac{\partial ll_{i}(\beta_{hu}^{(t)})}{\partial\beta_{hu}} \label{b-formula}\\
a_{hu}^{(t)} & =-\dfrac{1}{n}\sum_{i=1}^{n}\dfrac{\partial^{2}ll_{i}(\beta_{hu}^{(t)})}{\partial\beta_{hu}^{2}}. \label{a-formula}
\end{align}

Similar to \cite{friedman2010regularization}, $\beta_{hu}$ is then updated by minimizing
\begin{equation}
L_{1}^{(t)}(\beta_{hu})+d_{hu}^{(t)}\left|\beta_{hu}\right|+e_{hu}^{(t)}\beta_{hu}^{2}/2
\label{2nd-approx-LL-penalty}
\end{equation}
which has a closed form solution
\begin{equation}
\beta_{hu}^{(t+1)}=\dfrac{\mbox{sign}\left(a_{hu}^{(t)}\beta_{hu}^{(t)}-b_{hu}^{(t)}\right)}{a_{hu}^{(t)}+e_{hu}^{(t)}}\cdot\left(\left|a_{hu}^{(t)}\beta_{hu}^{(t)}-b_{hu}^{(t)}\right|-d_{hu}^{(t)}\right)_{+}
\label{update-beta}
\end{equation}
if $a_{hu}^{(t)} + e_{hu}^{(t)}>0$. From \eqref{e-formula}, \eqref{2nd-deri-beta} and \eqref{a-formula} we know that $a_{hu}^{(t)} \geq 0$ and $e_{hu}^{(t)} \geq 0$. 

The case $a_{hu}^{(t)} + e_{hu}^{(t)}=0$ implies that (i) $\alpha_{h}^{(t)}=\rho_{h}^{(t)}=\gamma_{h}^{(t)}=0$
 so that $\lambda_{h}^{(t)}(g)\equiv0$ (ii) $\beta_{hv}^{(t)}=0$ for $v \neq u$ or (iii) $\eta=1$ and $p_{i}^{(t)}=0$ or 1, $\forall i$. For the former two cases, the lower triangular part of the component matrix  $\lambda_{h}(g)^{(t)}\boldsymbol{\beta}_{h}^{(t)}\boldsymbol{\beta}_{h}^{(t)\top}$ becomes zero no matter what $\beta_{hu}$ takes. So we set $\beta_{hu}^{(t+1)}=0$ in these cases. Regarding (iii), if $p_i^{(t)} \equiv y_i$, $\forall i$, \eqref{2nd-approx-LL-penalty} is minimized at $\beta_{hu}=0$ because $b_{hu}^{(t)}=0$ at this time according to \eqref{1st-deri-beta} and \eqref{b-formula} together with $a_{hu}^{(t)} = e_{hu}^{(t)}=0$. Then the first derivative of \eqref{2nd-approx-LL-penalty} becomes $d_{hu}^{(t)}$ ($>0$) when $\beta_{hu}>0$ and $-d_{hu}^{(t)}$ ($<0$) when $\beta_{hu}<0$. Otherwise, $p_{i}^{(t)}=0$ or 1 $\forall i$ may be due to bad initialization. For example, a large magnitude of initial values of the parameters could easily make $p_i$ become 1 or 0, $\forall$ i through the logit link \eqref{SBLR-logit1}. In this case, setting $\beta_{hu}=0$ could prevent the divergence of the solution. 
In practice, we always normalize the entries in $\{W_{i}^{(s)}\}$ and the age terms $g_{is}$, $g_{is}^2$ as described at the beginning of Section \ref{estimation} before applying SBLR. Then we recommend to initialize each parameter from $U(-1/K, 1/K)$ to avoid the explosion in the logit scale. 

The computational complexity of updating each entry $\beta_{hu}$ is $O(nV)$ per iteration and thus that of updating $\left\{ \boldsymbol{\beta}_{h}\right\} _{h=1}^{K}$ is $O(nKV^2)$.

\subsection{Updates for $\{(\alpha_{h},\rho_{h},\gamma_{h}):h=1,\dots,K\}$}

Minimizing the loss function \eqref{loss-func} with respect to $\alpha_{h}$ when fixing the others is equivalent to:
\begin{equation}
\min_{\alpha_{h}}\ L(\alpha_{h})=-\dfrac{1}{n}\sum_{i=1}^{n}ll_{i}(\alpha_{h})+d_{h}\left|\alpha_{h}\right|+e_{h}\dfrac{\alpha_{h}^{2}}{2}
\label{optimize-alpha}
\end{equation}
where $d_{h}=\delta\eta\sum_{u=1}^{V}\sum_{v<u}\left|\beta_{hu}\right|\left|\beta_{hv}\right|$ and $e_{h}=\delta(1-\eta)\sum_{u=1}^{V}\sum_{v<u}\beta_{hu}^2 \beta_{hv}^2$.

Since $\alpha_h$ is linear in the logit link \eqref{SBLR-logit1} when fixing the other parameters, the log-likelihood $ll_i$ is a concave function of $\alpha_h$ given the others. Suppose the current estimate for $\alpha_h$ is $\alpha_h^{(t)}$. Similar to Section \ref{sec:update-beta}, $\alpha_h$ is updated by maximizing the following quadratic approximation to \eqref{optimize-alpha}:
\begin{equation}
c_{\alpha,h}^{(t)}+b_{\alpha,h}^{(t)}\left(\alpha_{h}-\alpha_{h}^{(t)}\right)+\dfrac{a_{\alpha,h}^{(t)}}{2}\left(\alpha_{h}-\alpha_{h}^{(t)}\right)^{2}+d_{h}^{(t)}\left|\alpha_{h}\right|+e_{h}^{(t)}\dfrac{\alpha_{h}^{2}}{2}
\label{2nd-approx-LL-alpha}
\end{equation}
where
\begin{equation}
\begin{aligned}
c_{\alpha,h}^{(t)} & =-\dfrac{1}{n}\sum_{i=1}^{n}ll_{i}(\alpha_{h}^{(t)})\\
b_{\alpha,h}^{(t)} & =-\dfrac{1}{n}\sum_{i=1}^{n}\dfrac{\partial ll_{i}(\alpha_{h}^{(t)})}{\partial\alpha_{h}}=-\dfrac{1}{n}\sum_{i=1}^{n}(y_{i}-p_{i}^{(t)})\boldsymbol{\beta}_{h}^{(t)\top}\left(\dfrac{1}{T_{i}}\sum_{s=1}^{T_{i}}\tilde{W}_{i}^{(s)}\right)\boldsymbol{\beta}_{h}^{(t)}\\
a_{\alpha,h}^{(t)} & =-\dfrac{1}{n}\sum_{i=1}^{n}\dfrac{\partial^{2}ll_{i}(\alpha_{h}^{(t)})}{\partial\alpha_{h}^{2}}=\dfrac{1}{n}\sum_{i=1}^{n}p_{i}^{(t)}(1-p_{i}^{(t)})\left[\boldsymbol{\beta}_{h}^{(t)\top}\left(\dfrac{1}{T_{i}}\sum_{s=1}^{T_{i}}\tilde{W}_{i}^{(s)}\right)\boldsymbol{\beta}_{h}^{(t)}\right]^{2}.
\end{aligned}
\label{2nd-approx-LL-coef-alpha}
\end{equation}
Note that $a_{\alpha,h}^{(t)} \geq 0$ and $e_{h}^{(t)} \geq 0$. If $a_{\alpha,h}^{(t)} + e_{h}^{(t)} > 0$, $\alpha_h$ is then updated to the argmin of \eqref{2nd-approx-LL-alpha}:
\begin{equation}
\alpha_{h}^{(t+1)}=\dfrac{\mbox{sign}\left(a_{\alpha,h}^{(t)}\alpha_{h}^{(t)}-b_{\alpha,h}^{(t)}\right)}{a_{\alpha,h}^{(t)}+e_{h}^{(t)}}\left(\left|a_{\alpha,h}^{(t)}\alpha_{h}^{(t)}-b_{\alpha,h}^{(t)}\right|-d_{h}^{(t)}\right)_{+}.
\label{update-alpha}
\end{equation}
If $a_{\alpha,h}^{(t)} + e_{h}^{(t)} = 0$, then either the component matrix $\boldsymbol{\beta}_{h}^{(t)}\boldsymbol{\beta}_{h}^{(t)\top}$ is a zero matrix or $p_{i}^{(t)}=0$ or 1, $\forall i$. In either case, $\alpha_h^{(t+1)}$ is set to 0 following a similar discussion in Section \ref{sec:update-beta}.

Likewise, the update rules for each $\rho_h$ and $\gamma_h$ are as follows.
\begin{equation}
\rho_{h}^{(t+1)}=\begin{cases}
\dfrac{\mbox{sign}\left(a_{\rho,h}^{(t)}\rho_{h}^{(t)}-b_{\rho,h}^{(t)}\right)}{a_{\rho,h}^{(t)}+e_{h}^{(t)}}\left(\left|a_{\rho,h}^{(t)}\rho_{h}^{(t)}-b_{\rho,h}^{(t)}\right|-d_{h}^{(t)}\right)_{+} & \mbox{if }a_{\rho,h}^{(t)} + e_h^{(t)}>0\\
0 & \mbox{otherwise}
\end{cases}
\label{update-rho}
\end{equation}
where
\begin{equation}
\begin{aligned}
b_{\rho,h}^{(t)} & =-\dfrac{1}{n}\sum_{i=1}^{n}\dfrac{\partial ll_{i}(\rho_{h}^{(t)})}{\partial\rho_{h}}=-\dfrac{1}{n}\sum_{i=1}^{n}(y_{i}-p_{i}^{(t)})\boldsymbol{\beta}_{h}^{(t)\top}\left(\dfrac{1}{T_{i}}\sum_{s=1}^{T_{i}}\tilde{g}_{is}\tilde{W}_{i}^{(s)}\right)\boldsymbol{\beta}_{h}^{(t)} \\
a_{\rho,h}^{(t)} & =-\dfrac{1}{n}\sum_{i=1}^{n}\dfrac{\partial^{2}ll_{i}(\rho_{h}^{(t)})}{\partial\rho_{h}^{2}}=\dfrac{1}{n}\sum_{i=1}^{n}p_{i}^{(t)}(1-p_{i}^{(t)})\left[\boldsymbol{\beta}_{h}^{(t)\top}\left(\dfrac{1}{T_{i}}\sum_{s=1}^{T_{i}}\tilde{g}_{is}\tilde{W}_{i}^{(s)}\right)\boldsymbol{\beta}_{h}^{(t)}\right]^{2}. 
\end{aligned}
\label{2nd-approx-LL-coef-rho}
\end{equation}
And
\begin{equation}
\gamma_{h}^{(t+1)}=\begin{cases}
\dfrac{\mbox{sign}\left(a_{\gamma,h}^{(t)}\gamma_{h}^{(t)}-b_{\gamma,h}^{(t)}\right)}{a_{\gamma,h}^{(t)}+e_{h}^{(t)}}\left(\left|a_{\gamma,h}^{(t)}\gamma_{h}^{(t)}-b_{\gamma,h}^{(t)}\right|-d_{h}^{(t)}\right)_{+} & \mbox{if }a_{\gamma,h}^{(t)} + e_h^{(t)}>0\\
0 & \mbox{otherwise}
\end{cases}
\label{update-gamma}
\end{equation}
where
\begin{equation}
\begin{aligned}
b_{\gamma,h}^{(t)} & =-\dfrac{1}{n}\sum_{i=1}^{n}\dfrac{\partial ll_{i}(\gamma_{h}^{(t)})}{\partial\gamma_{h}}=-\dfrac{1}{n}\sum_{i=1}^{n}(y_{i}-p_{i}^{(t)})\boldsymbol{\beta}_{h}^{(t)\top}\left(\dfrac{1}{T_{i}}\sum_{s=1}^{T_{i}}\widetilde{g_{is}^{2}}\tilde{W}_{i}^{(s)}\right)\boldsymbol{\beta}_{h}^{(t)} \\
a_{\gamma,h}^{(t)} & =-\dfrac{1}{n}\sum_{i=1}^{n}\dfrac{\partial^{2}ll_{i}(\gamma_{h}^{(t)})}{\partial\gamma_{h}^{2}}=\dfrac{1}{n}\sum_{i=1}^{n}p_{i}^{(t)}(1-p_{i}^{(t)})\left[\boldsymbol{\beta}_{h}^{(t)\top}\left(\dfrac{1}{T_{i}}\sum_{s=1}^{T_{i}}\widetilde{g_{is}^{2}}\tilde{W}_{i}^{(s)}\right)\boldsymbol{\beta}_{h}^{(t)}\right]^{2} .
\end{aligned}
\label{2nd-approx-LL-coef-gamma}
\end{equation}

The computational complexity for updating each $\alpha_h$, $\rho_h$ or $\gamma_h$ is $O(nV^2)$ per iteration and hence that of updating $\{(\alpha_{h},\rho_{h},\gamma_{h})\}_{h=1}^K$ is $O(nKV^2)$. This step requires storing the transformed matrix predictors for each subject, $\left\{ \sum_{s=1}^{T_{i}}\tilde{W}_{i}^{(s)}/T_i\right\} _{i=1}^{n}$, $\left\{ \sum_{s=1}^{T_{i}}\tilde{g}_{is}\tilde{W}_{i}^{(s)}/T_i\right\} _{i=1}^{n}$, $\left\{ \sum_{s=1}^{T_{i}}\widetilde{g_{is}^{2}}\tilde{W}_{i}^{(s)}/T_i\right\} _{i=1}^{n}$, \\ and the intermediate results $\left\{ \boldsymbol{\beta}_{h}^{(t)\top}\left(\sum_{s=1}^{T_{i}}\tilde{W}_{i}^{(s)}/T_{i}\right)\boldsymbol{\beta}_{h}^{(t)} \right\}_{h=1}^K$, \\ $\left\{ \boldsymbol{\beta}_{h}^{(t)\top}\left(\sum_{s=1}^{T_{i}}\tilde{g}_{is}\tilde{W}_{i}^{(s)}/T_i\right)\boldsymbol{\beta}_{h}^{(t)}\right\} _{h=1}^{K}$ and $\left\{ \boldsymbol{\beta}_{h}^{(t)\top}\left(\sum_{s=1}^{T_{i}}\widetilde{g_{is}^{2}}\tilde{W}_{i}^{(s)}/T_i\right)\boldsymbol{\beta}_{h}^{(t)}\right\} _{h=1}^{K}$ for each subject. Therefore the memory complexity is $O(nV^2+nK)$.

\subsection{Update for $\alpha_0$}

The intercept $\alpha_0$ is also updated by maximizing the quadratic approximation to the joint log-likelihood at the current estimate $\alpha_0^{(t)}$ with the updating rule
\begin{equation}
\alpha_{0}^{(t+1)}=\begin{cases}
\alpha_{0}^{(t)}-\dfrac{b^{(t)}}{a^{(t)}} & \mbox{if }a^{(t)}>0\\
0 & \mbox{otherwise}
\end{cases}
\label{update-intercept}
\end{equation}
where 
\begin{equation}
\begin{aligned}
b^{(t)} & =-\dfrac{1}{n}\sum_{i=1}^{n}\dfrac{\partial ll_{i}(\alpha_{0}^{(t)})}{\partial\alpha_{0}}=-\dfrac{1}{n}\sum_{i=1}^{n}(y_{i}-p_{i}^{(t)}) \\
a^{(t)} & =-\dfrac{1}{n}\sum_{i=1}^{n}\dfrac{\partial^{2}ll_{i}(\alpha_{0}^{(t)})}{\partial\alpha_{0}^{2}}=\dfrac{1}{n}\sum_{i=1}^{n}p_{i}^{(t)}(1-p_{i}^{(t)}).
\end{aligned}
\label{2nd-approx-LL-coef-intercept}
\end{equation}
The computational complexity of this step is $O(n)$.

\subsection{Other Details}
\label{other details}

The above procedure is cycled through all the parameters until the relative change of the loss function \eqref{loss-func} is smaller than a tolerance number $\epsilon$, as summerized in Algorithm \ref{CD-SBLR}. We set $\epsilon=10^{-5}$ in simulations and applications. Since the loss function \eqref{loss-func} is lower bounded by 0 and each update always decreases the function value, the coordinate descent algorithm derived above is guaranteed to converge.

\begin{algorithm}[t]
\caption{Coordinate descent for SBLR model with elastic-net penalty. \label{CD-SBLR}}

\begin{algorithmic}[1]

\State \textbf{Input:} Normalized $V\times V$ symmetric matrix observations $\{\tilde{W}_{i}^{(s)}:s=1,\dots,T_{i}\}$, standardized age $\{\tilde{g}_{is}\}$ and squared age $\{\widetilde{g_{is}^{2}}\}$, binary outcome $y_{i}$, $i=1,\dots,n$; rank $K$, overall penalty factor $\delta > 0$, $L_1$ fractional penalty factor $\eta \in (0,1]$, tolerance $\epsilon > 0$.

\State \textbf{Output:} Estimates of $\alpha_0$, $\left\{ \left(\alpha_{h},\rho_{h},\gamma_{h},\boldsymbol{\beta}_{h}\right):h=1,\dots,K\right\}$.

\State \textbf{Initialize:} $\alpha_0$ and each parameter of $\left\{ \left(\alpha_{h},\rho_{h},\gamma_{h},\boldsymbol{\beta}_{h}\right):h=1,\dots,K\right\} \sim U(-1/K, 1/K)$.

\Repeat
	\For{$h=1:K$}
		\For{$u=1:V$}
			\State update $\beta_{hu}$ by \eqref{update-beta}
		\EndFor
	\EndFor
	\For{$h=1:K$}
		\State update $\alpha_{h}$ by \eqref{update-alpha}
		\State update $\rho_h$ by \eqref{update-rho}
		\State update $\gamma_h$ by \eqref{update-gamma}
	\EndFor
	\State update $\alpha_0$ by \eqref{update-intercept}
\Until{relative change of loss function \eqref{loss-func} $<\epsilon$.}

\end{algorithmic}
\end{algorithm}

In general, the algorithm should be run from multiple initializations to locate a good local solution. Although the estimates for $\left\{ \boldsymbol{\beta}_{h}\right\} _{h=1}^{K}$ and $\left\{ \left(\alpha_{h},\rho_{h},\gamma_{h}\right)\right\} _{h=1}^{K}$ will all be zero under sufficiently large penalty factor $\delta$, we cannot initialize them at zero because the results will then get stuck at zero. The updating rules \eqref{update-beta} - \eqref{update-gamma} imply that each parameter will be set to 0 given others being zero. In fact, we recommend to initialize all the parameters nonzero in case some components get degenerated unexpectedly at the beginning. In practice, we initialize each parameter from a uniform distribution $U(-1/K,1/K)$ as discussed at the end of Section \ref{sec:update-beta}.

\subsection{Model Selection}

The penalty factors ($\delta$, $\eta$) in the regularization \eqref{elanet-pen} can be tuned by cross validation (CV) over a grid of values on $[\delta_{min},\delta_{max}]\times (0,1]$, where $\delta_{max}$ is a \emph{roughly} smallest value for which all the parameters $\left\{ \left(\boldsymbol{\beta}_{h}, \alpha_{h},\rho_{h},\gamma_{h}\right)\right\} _{h=1}^{K}$ become zero when $\eta$ takes the the smallest value, and $\delta_{min}$ is a sufficiently small value that produces dense results when $\eta=1$ (lasso). We record the deviance (minus twice the average log-likelihood) for each left-out fold rather than AUC or misclassification error, since deviance is smoother \citep{friedman2010regularization}. Then for each value of $\eta$, the optimal $\delta$ is collected such that the mean CV deviance is within one standard-error of the minimum. Among these pairs of $(\delta,\eta)$, we pick the one that produces the smallest deviance. This strategy is adapted from the ``one-standard-error" rule for lasso \citep{friedman2010regularization} that prefers a more parsimonious model since the risk curves are estimated with errors. We also refer to it as the ``one-standard-error" rule for selecting the best model under elastic-net penalty in the latter part of this paper.

Our proposed model \eqref{SBLR-model} assumes a known number of components $K$. In practice, we choose a large enough number for $K$ and allow the elastic-net penalty \eqref{elanet-pen} to discard unnecessary components, leading to a data-driven estimate for the number of signal subgraphs. We assess the performance of our procedure and verify its lack of sensitivity to the chosen upper bound $K$ in Table \ref{all-in-one table}.

\section{Simulation Study}

In this section, we first conduct a number of simulation experiments to evaluate the empirical computational and memory complexity of the coordinate descent algorithm derived in Section \ref{estimation}. We then compare the inference results to competitors.

\subsection{Computational and Memory Complexity}

\begin{figure}[t]
	\centering
	\includegraphics[width=.45\textwidth]{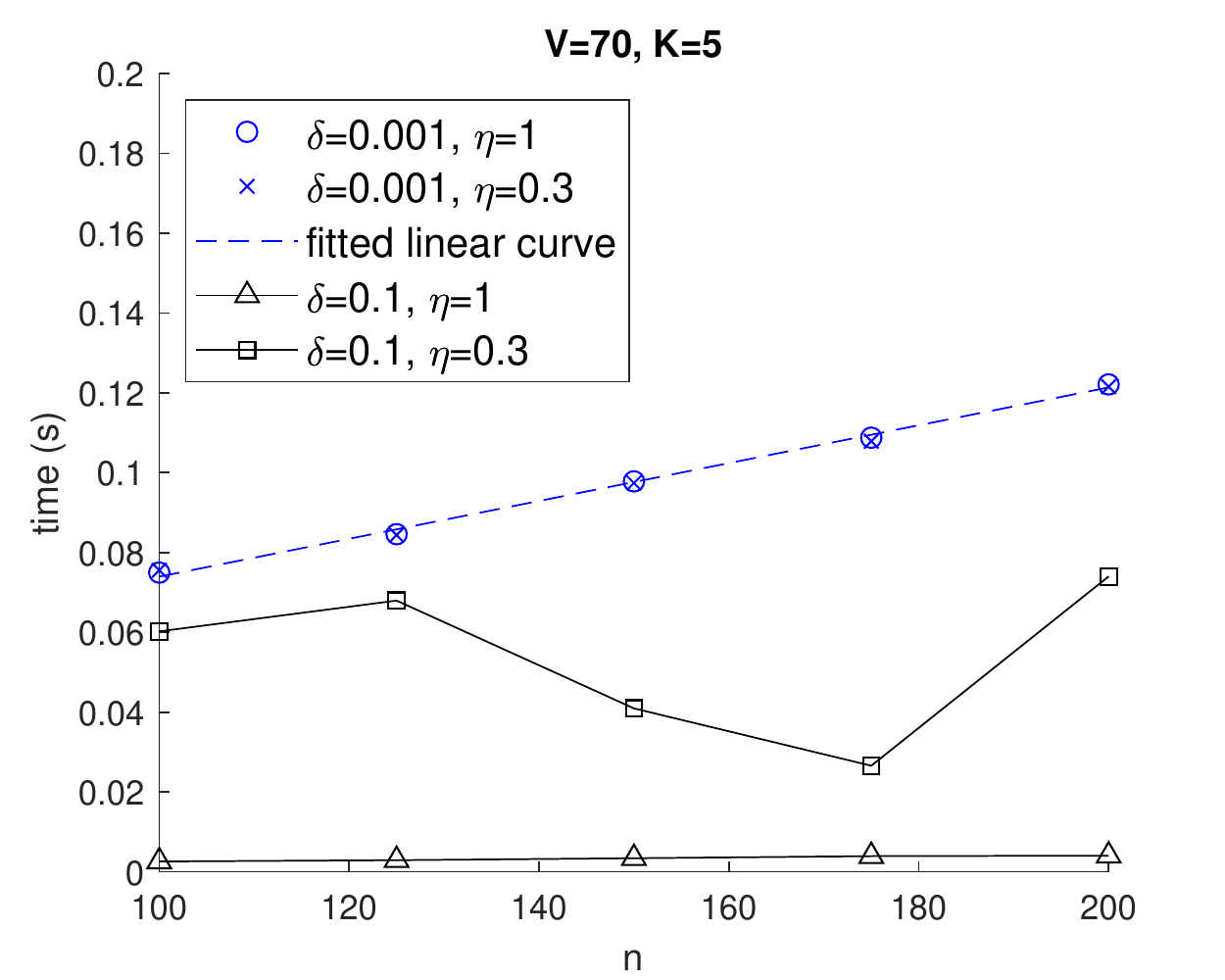} \includegraphics[width=.45\textwidth]{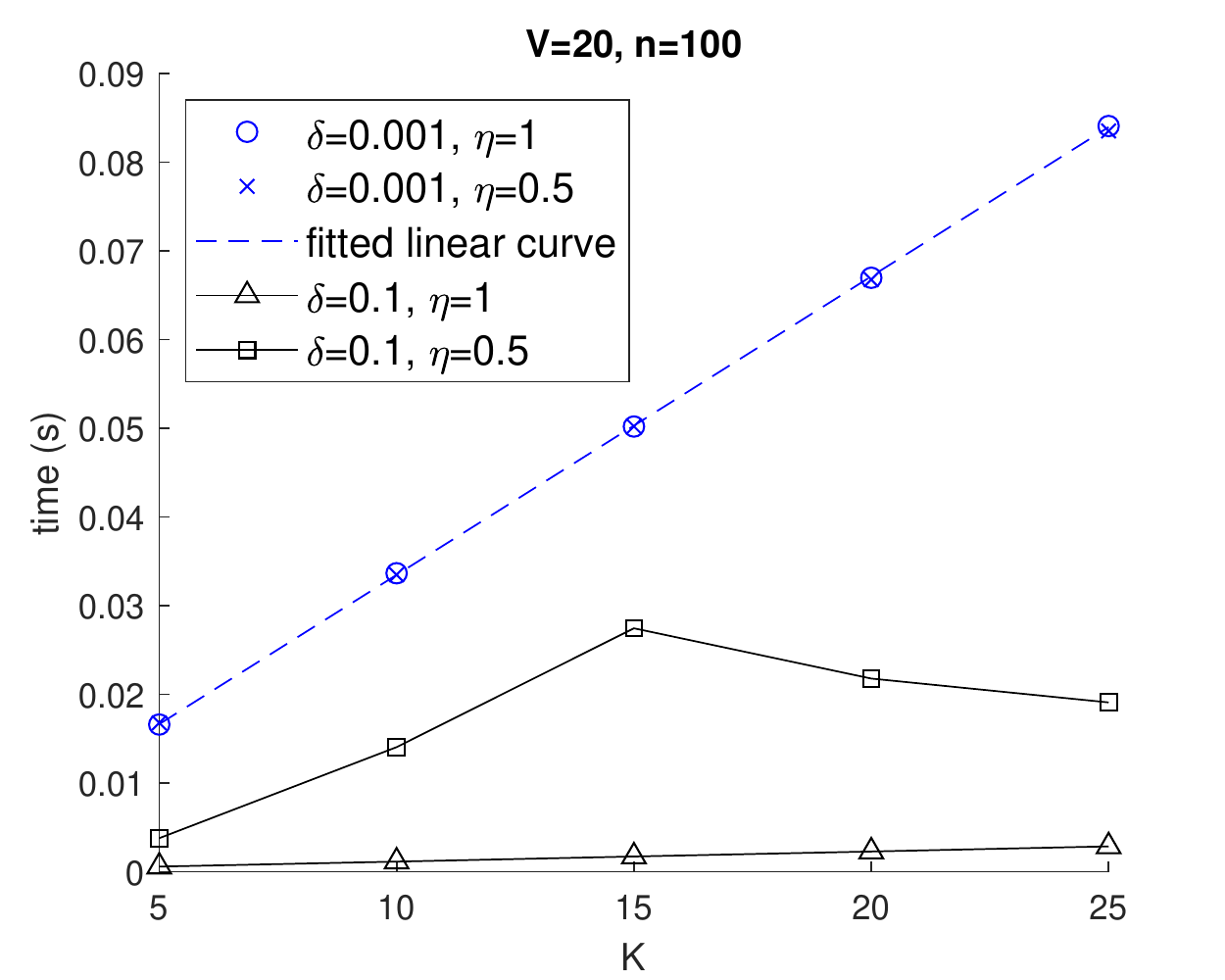}
	\caption{Average computational time (in seconds) per iteration for 30 runs versus the number of subjects $n$ (left) and the number of components $K$ (right).}
	\label{comp-n-K}
\end{figure}

\begin{figure}[t]
	\centering
	\includegraphics[width=.45\textwidth]{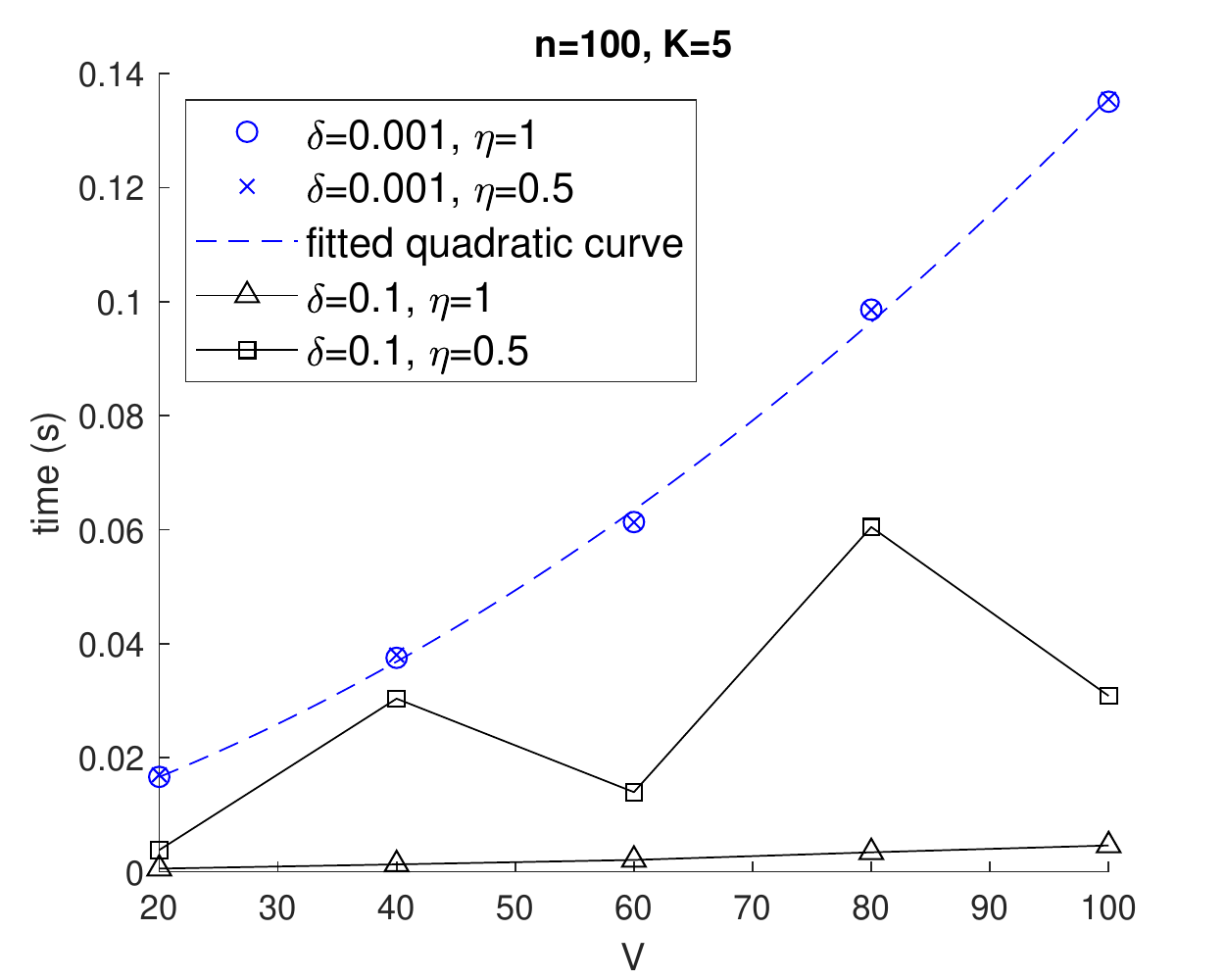} \includegraphics[width=.45\textwidth]{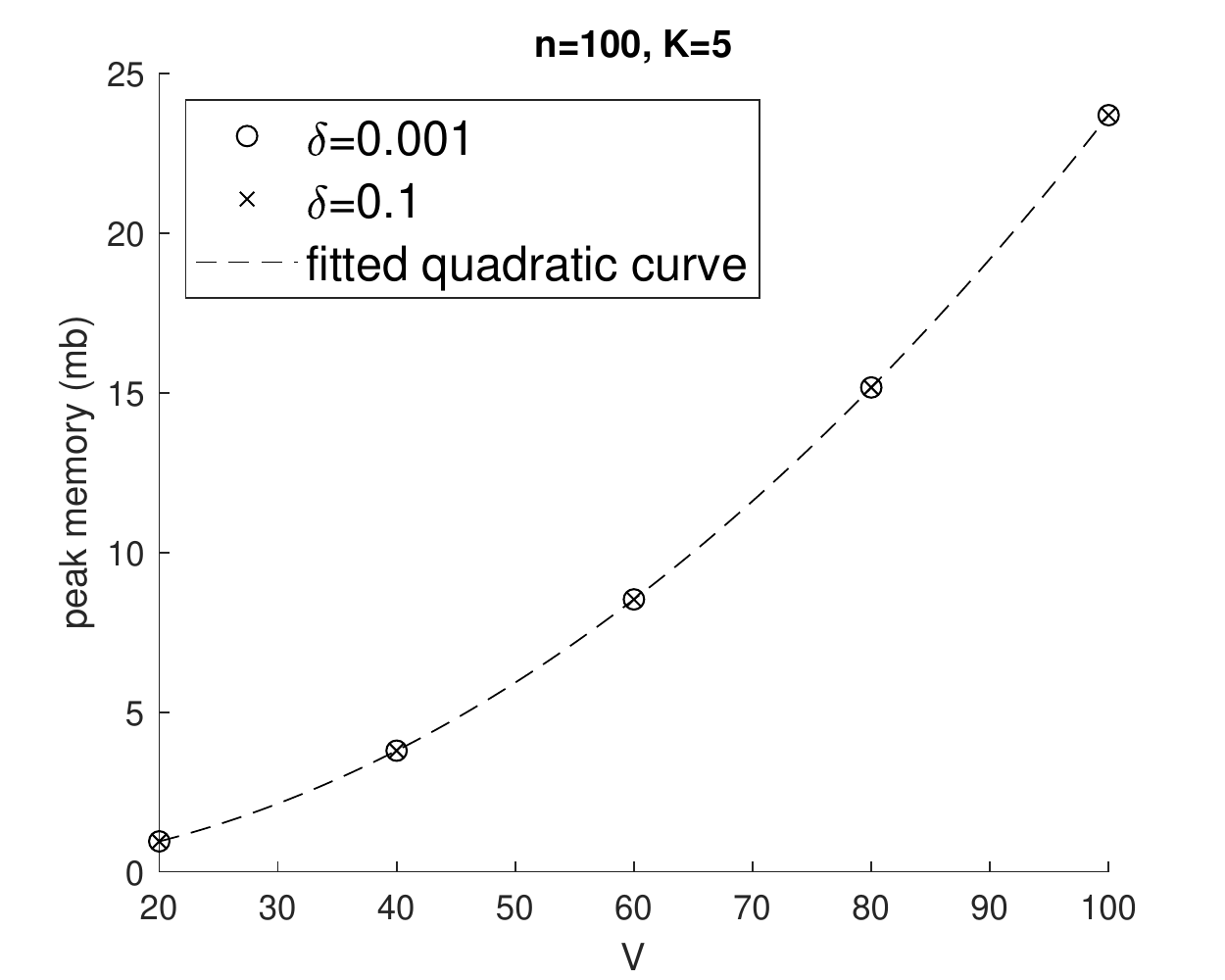}
	\caption{Average computational time (in seconds) per iteration (left) and average peak memory (in mb) in use (right) for 30 runs versus the number of nodes $V$.}
	\label{comp-mem-V}
\end{figure}

Algorithm \ref{CD-SBLR} is implemented in Matlab (R2018a) and the code is publicly available in Github (\url{https://github.com/wangronglu/Symmetric-Bilinear-Logistic-Regression}). All the numerical experiments are conducted in a machine with six Intel Core i7 3.2 GHz processors and 64 GB of RAM. We simulated observations $\left(y_{i},\{g_{is},W_{i}^{(s)}:s=1,\dots,T_{i}\}\right)$, $i=1,\dots,n$ for different number of subjects $n$ and different number of nodes $V$, where $W_{i[uv]}^{(s)}=W_{i[vu]}^{(s)}\sim N(0,1)\ (u\neq v)$, $g_{i1}\sim U(60,90)$, $T_{i}\sim U\{1,2,\dots,5\}$, $y_{i}\sim$ Bernoulli(0.5). Figure \ref{comp-n-K} and \ref{comp-mem-V} display how the execution time and peak memory (maximum amount of memory in use) vary with different problem sizes. 

Note that the computational time of Algorithm \ref{CD-SBLR} also depends on the penalty factors $\delta$ and $\eta$, because smaller $\delta$ and $\eta$ will lead to more nonzero estimates for the parameters. Figure \ref{comp-n-K} and the left plot of Figure \ref{comp-mem-V} show that the runtime per iteration is a linear order with $n$ and $K$, and a quadratic order with $V$ under small $\delta$ and $\eta$, which is in accordance with our analysis of computational complexity in Section \ref{estimation}, i.e. $O(nKV^2)$ in the worst case scenario.

The right plot of Figure \ref{comp-mem-V} shows that the peak memory of our algorithm is a quadratic order with $V$, no matter what penalty factors are used. This is in accordance with our analysis of memory complexity in Section \ref{estimation}, which is $O(nV^2+nK)$ and dominated by the quadratic term of $V$ in these cases.

Our algorithm is coded in the Matlab programming environment using no \texttt{C} or \texttt{FORTRAN} code. There are substantial margins to reduce the computational time if such code were used, as each iteration involves for-loops over the entries of component vectors $\left\{ \boldsymbol{\beta}_{h}\right\} _{h=1}^{K}$, which are particularly slow in Matlab.

\subsection{Inference on Signal Subgraphs and Age Effects}
\label{sec:simu2}

In this experiment, we compare the performance of recovering true signal subgraphs and age effects among SBLR model and the following methods:
\begin{enumerate}
	\item Unstructured logistic regression (LR) with elastic-net penalty. This model does not assume any structure on the coefficient matrices:
	\begin{equation}
	\mbox{logit}(p_{i})=\alpha_{0}+\left\langle B_{1},\dfrac{1}{T_{i}}\sum_{s=1}^{T_{i}}W_{i}^{(s)}\right\rangle +\left\langle B_{2},\dfrac{1}{T_{i}}\sum_{s=1}^{T_{i}}g_{is}W_{i}^{(s)}\right\rangle +\left\langle B_{3},\dfrac{1}{T_{i}}\sum_{s=1}^{T_{i}}g_{is}^{2}W_{i}^{(s)}\right\rangle
	\label{unstructure-LR}
	\end{equation} 
	where $B_1$, $B_2$ and $B_3$ are $V \times V$ symmetric coefficient matrices. In fact we only need to enter the lower triangular entries of $\dfrac{1}{T_{i}}\sum_{s=1}^{T_{i}}W_{i}^{(s)}$, $\dfrac{1}{T_{i}}\sum_{s=1}^{T_{i}}g_{is}W_{i}^{(s)}$ and $\dfrac{1}{T_{i}}\sum_{s=1}^{T_{i}}g_{is}^{2}W_{i}^{(s)}$ in estimating \eqref{unstructure-LR}. We write \eqref{unstructure-LR} in matrix dot product form for concision and convenience of displaying results. 
	Elastic-net penalty is put on the entries of the coefficient matrices $B_1$, $B_2$ and $B_3$ in model estimation. 	
	
	\item Naively symmetrized tensor regression (NSTR). According to \cite{zhou2013tensor}, a generic rank-$K$ tensor regression for binary response on the matrix predictors in \eqref{SBLR-logit1} has the form
	\begin{equation}
	\mbox{logit}(p_{i})=\alpha_{0}+\sum_{h=1}^{K}\left\langle \boldsymbol{\beta}_{1}^{(h)}\circ\boldsymbol{\beta}_{2}^{(h)}\circ\boldsymbol{\beta}_{3}^{(h)},\boldsymbol{X}_{i}\right\rangle 
	\label{naive-tr}
	\end{equation}
	where $\circ$ denotes outer product, $\left\langle \cdot,\cdot\right\rangle$ denotes tensor dot product \citep{kolda2009tensor}, and $\boldsymbol{\beta}_{1}^{(h)}\in\mathbb{R}^{V}$, $\boldsymbol{\beta}_{2}^{(h)}\in\mathbb{R}^{V}$,
	$\boldsymbol{\beta}_{3}^{(h)}\in\mathbb{R}^{3}$, $h=1,\dots,K$. $\boldsymbol{X}_{i}$ in \eqref{naive-tr} is a $V\times V\times 3$ array constructed by concatenating the matrices $\dfrac{1}{T_{i}}\sum_{s=1}^{T_{i}}W_{i}^{(s)}$, $\dfrac{1}{T_{i}}\sum_{s=1}^{T_{i}}g_{is}W_{i}^{(s)}$ and $\dfrac{1}{T_{i}}\sum_{s=1}^{T_{i}}g_{is}^{2}W_{i}^{(s)}$ for subject $i$ as shown in Figure \ref{tensor-reg}. 
	
	\begin{figure}[ht]
		\centering
		\includegraphics[width=.7\textwidth]{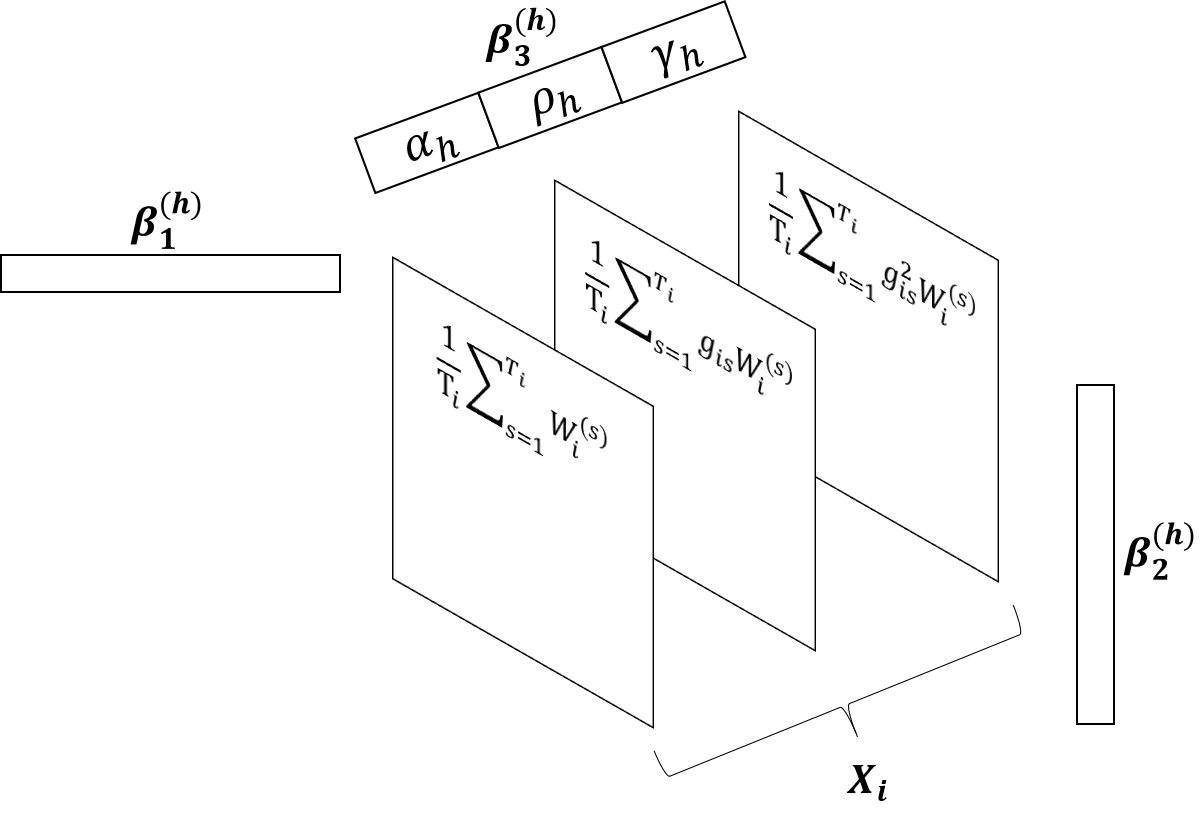}
		\caption{The 3D array $\boldsymbol{X}_{i}$ constructed for each subject $i$ in a tensor regression.}
		\label{tensor-reg}
	\end{figure}

	Then the nonzero entries in each component matrix $\boldsymbol{\beta}_{1}^{(h)}\circ\boldsymbol{\beta}_{2}^{(h)}=\boldsymbol{\beta}_{1}^{(h)}\boldsymbol{\beta}_{2}^{(h)\top}$ are supposed to locate an outcome-relevant subgraph. However, the partial symmetry in $\boldsymbol{X}_{i}$ when fixing the 3rd dimension does not necessarily lead to the same estimates for $\boldsymbol{\beta}_{1}^{(h)}$ and $\boldsymbol{\beta}_{2}^{(h)}$. A naive solution is to use the symmetrized component matrices
	\begin{equation}
	\left\{ (\boldsymbol{\beta}_{1}^{(h)}\boldsymbol{\beta}_{2}^{(h)\top}+\boldsymbol{\beta}_{2}^{(h)}\boldsymbol{\beta}_{1}^{(h)\top})/2:h=1,\dots,K\right\}  
	\label{naive-est}
	\end{equation} 
	to locate signal subgraphs. Let $\boldsymbol{\beta}_{3}^{(h)}=(\alpha_{h},\rho_{h},\gamma_{h})^{\top}$. Then the age effect of the $h$-th signal subgraph in \eqref{naive-tr} is estimated as $\lambda_{h}(g)=\gamma_{h}g^{2}+\rho_{h}g+\alpha_{h}$.
	According to \cite{zhou2013tensor}, we add the following elastic-net penalty on component vectors in \eqref{naive-tr} to encourage sparsity:
	\begin{equation}
	\begin{aligned}
	& \rho\sum_{h=1}^{K}\left[\dfrac{(\lambda-1)}{2}\left(\left\Vert \boldsymbol{\beta}_{1}^{(h)}\right\Vert _{2}^{2}+\left\Vert \boldsymbol{\beta}_{2}^{(h)}\right\Vert _{2}^{2}+\left\Vert \boldsymbol{\beta}_{3}^{(h)}\right\Vert _{2}^{2}\right)+\right.\\
	& \qquad\quad\left.(2-\lambda)\left(\left\Vert \boldsymbol{\beta}_{1}^{(h)}\right\Vert _{1}+\left\Vert \boldsymbol{\beta}_{2}^{(h)}\right\Vert _{1}+\left\Vert \boldsymbol{\beta}_{3}^{(h)}\right\Vert _{1}\right)\right],\ \lambda\in[1,2].
	\end{aligned}
	\label{tensor-reg-penalty}
	\end{equation}
	
\end{enumerate}

\begin{figure}[ht]
	\centering
	\includegraphics[width=.5\textwidth]{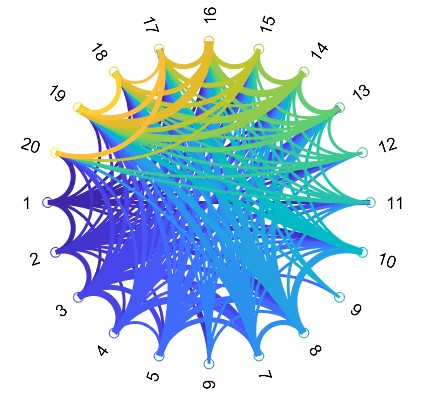} 
	\caption{Overlay of 11 basis subgraphs corresponding to $\left\{ \boldsymbol{q}_{h}\boldsymbol{q}_{h}^{\top}\right\} _{h=1}^{11}$.}
	\label{basis-sup}
\end{figure}

We simulate a synthetic dataset of 100 subjects. To mimic the real data, the number of network observations for each subject ranges from 1 to 5 randomly and the age varies over a 5-year span accordingly with $g_{i1}\sim U(60,90)$, $i=1,\dots, 100$. Each network has 20 nodes and the corresponding adjacency matrices $\{W_{i}^{(s)}\}$ are generated as follows. The first network observation $W_i^{(1)}$ for each subject is generated from a set of basis subgraphs to induce correlations among edges with an individual loading vector as
\begin{equation}
W_{i}^{(1)}=\sum_{h=1}^{11}\lambda_{ih}\boldsymbol{q}_{h}\boldsymbol{q}_{h}+\Delta_{i}
\label{generate-networks}
\end{equation}
where $\boldsymbol{q}_{h}\in\{0,1\}^{20}$ is a random binary vector with $\left\Vert \boldsymbol{q}_{h}\right\Vert _{0} = h+1$, $h=1,\dots, 10$ and $\left\Vert \boldsymbol{q}_{11}\right\Vert _{0}=4$. The loadings $\{\lambda_{ih}\}$ in \eqref{generate-networks} are generated independently from $U(0,1)$ and $\Delta_i$ adds 5\% random noise to each connection strength in the network. Figure \ref{basis-sup} visualizes the 11 basis subgraphs superimposed together, showing that the generating process \eqref{generate-networks} produces networks with complex correlation structure. The follow-up network observation $W_i^{(s)}$ for each subject is generated by adding $N(0, 1)$ percent of relative change to each connection strength in $W_i^{(s-1)}$, $s=2,\dots, T_i$. Note that $T_i \leq T_m = 5$, $\forall i$.

\begin{figure}[!tb]
	\centering
	\includegraphics[width=.32\textwidth]{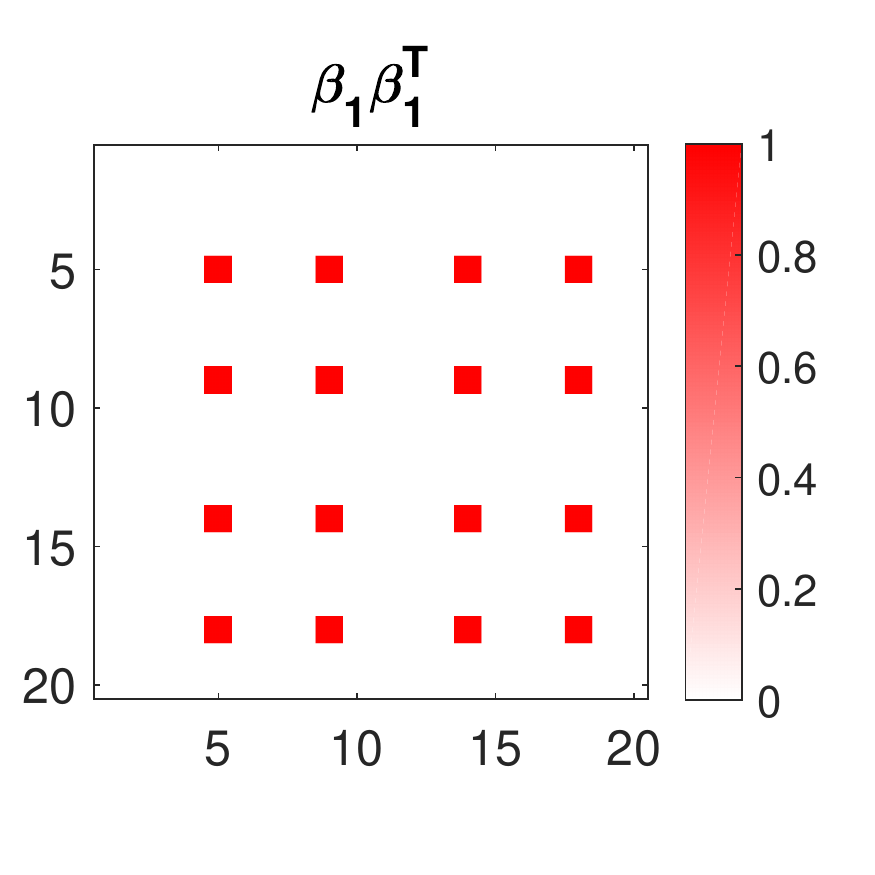}\hspace{5pt}\includegraphics[width=.29\textwidth]{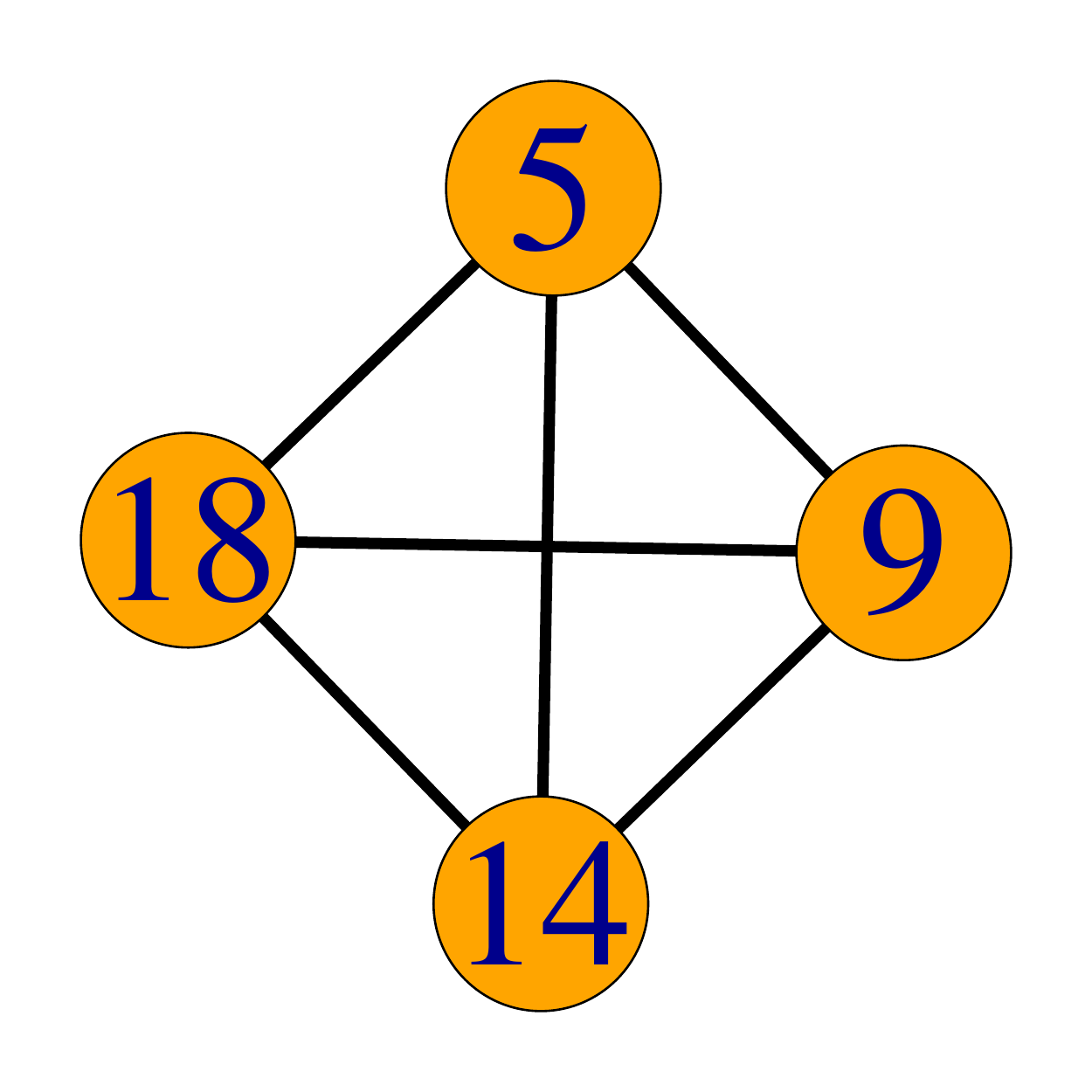}\hspace{5pt}\includegraphics[width=.31\textwidth]{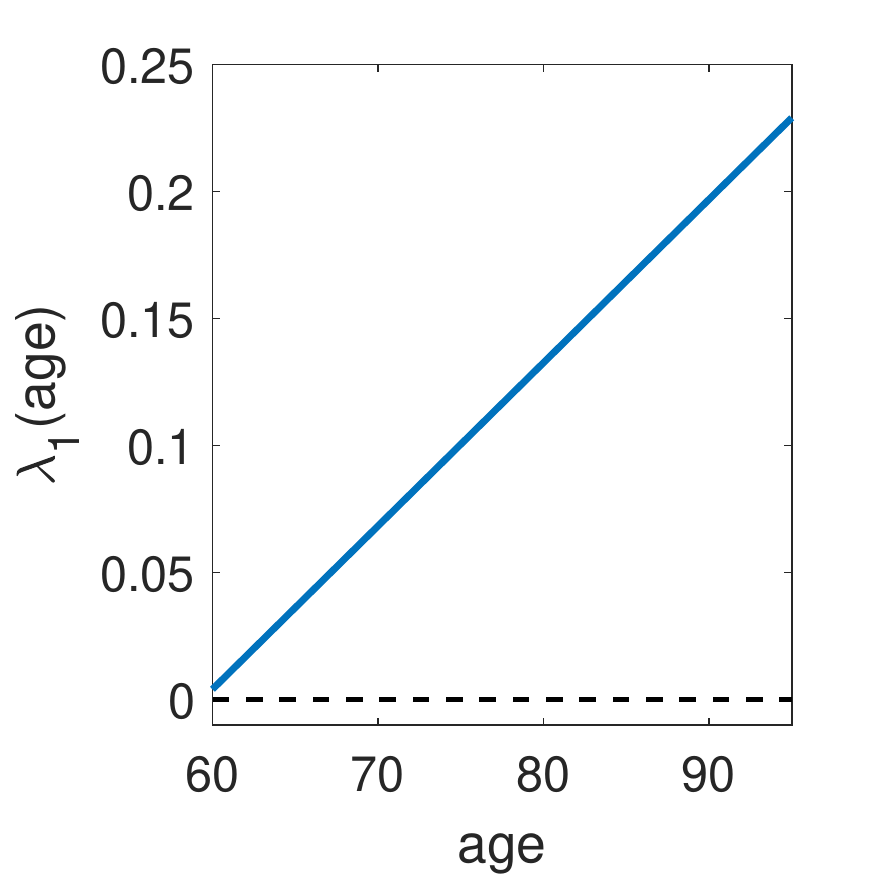} \\
	\includegraphics[width=.32\textwidth]{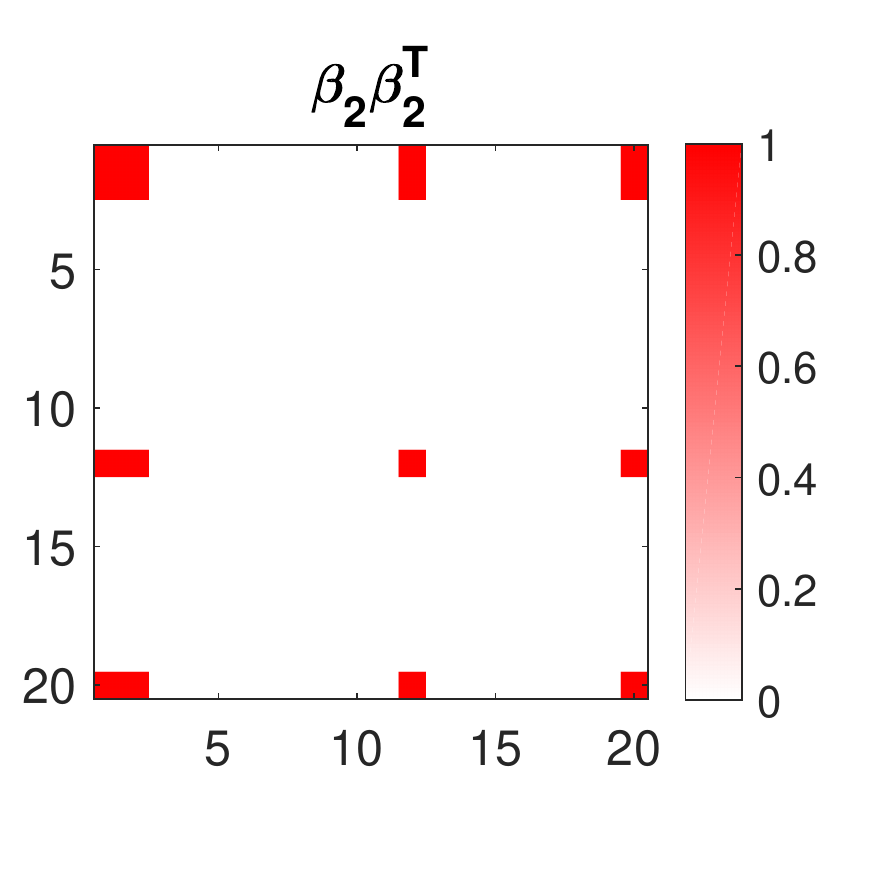}\hspace{5pt}\includegraphics[width=.29\textwidth]{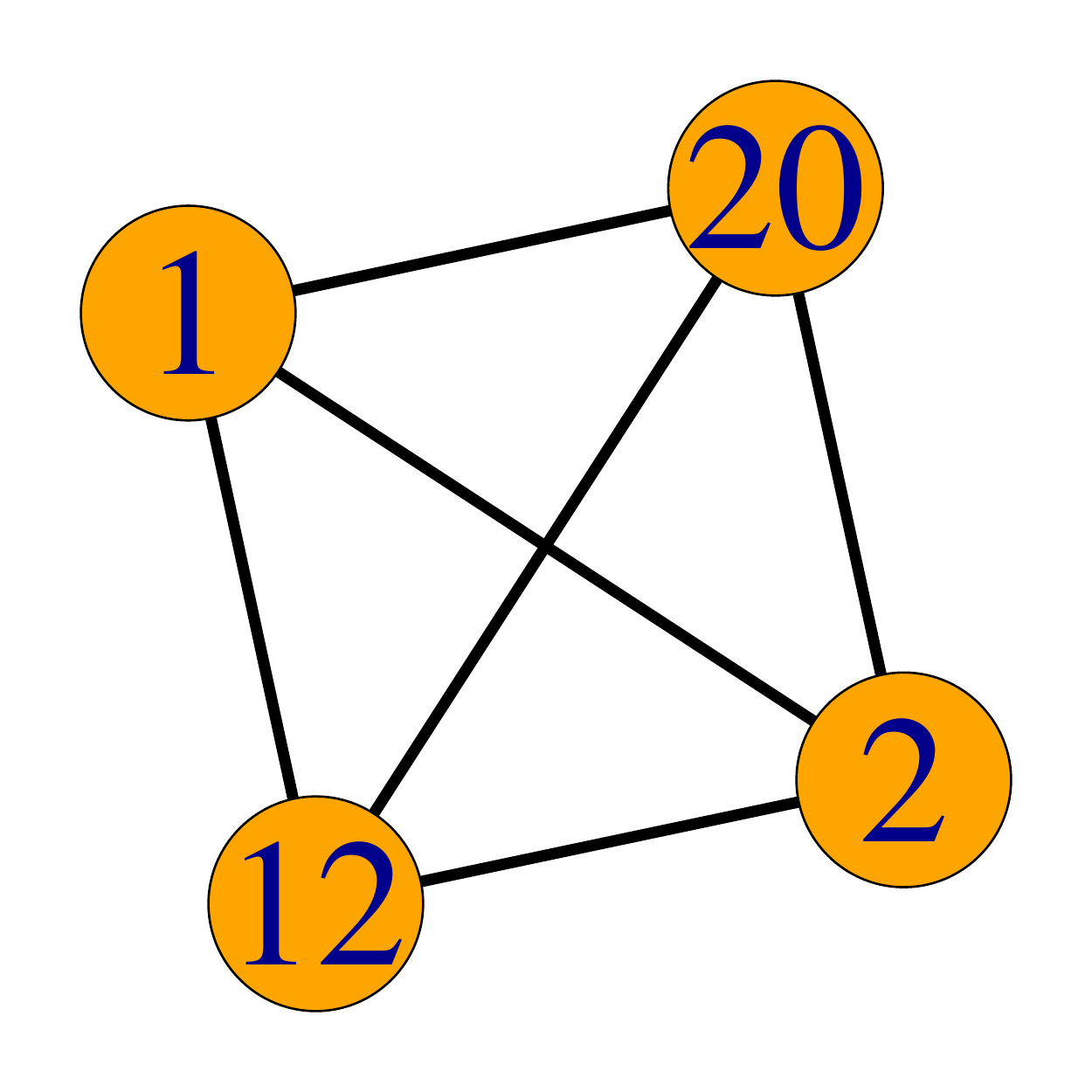}\hspace{5pt}\includegraphics[width=.31\textwidth]{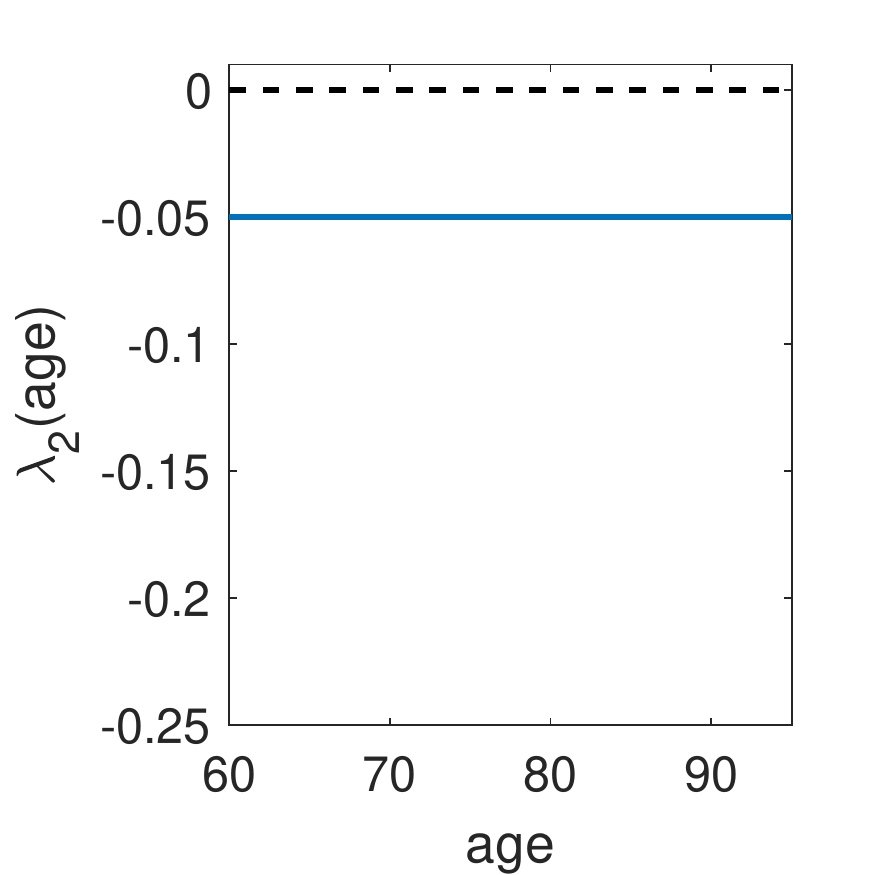}
	\caption{True coefficient matrix $\boldsymbol{\beta}_{h}\boldsymbol{\beta}_{h}^{\top}$ for $h=1$ (upper) and $h=2$ (lower) in a simulated dataset, along with the corresponding signal subgraphs (middle) and age effects $\{\lambda_{h}(g)\}$ (right).}
	\label{true-signal-subgraphs}
\end{figure}

The binary response $y_i$ is generated from Bernoulli($p_i$) independently with 
\begin{equation}
\mbox{logit}(p_{i})=\sum_{h=1}^{2}\dfrac{1}{T_{i}}\sum_{s=1}^{T_{i}}\lambda_{h}(g_{is})\boldsymbol{\beta}_{h}^{\top}\tilde{W}_{i}^{(s)}\boldsymbol{\beta}_{h}
\label{generate-response}
\end{equation}
where $\boldsymbol{\beta}_{1}=\boldsymbol{q}_{3}$, $\boldsymbol{\beta}_{2}=\boldsymbol{q}_{11}$ in \eqref{generate-networks}, and the functions
$\{\lambda_{h}(g):h=1,2\}$ are set as the right column in Figure \ref{true-signal-subgraphs}. The generating process \eqref{generate-networks} - \eqref{generate-response} indicate that the true signal subgraphs relevant to $y$ correspond to two basis subgraphs $\{\boldsymbol{q}_{h}\boldsymbol{q}_{h}^{\top}:h=3,11\}$, each with 4 nodes and linear or constant age effect as displayed in Figure \ref{true-signal-subgraphs}.

\begin{figure}[htb]
\centering
\includegraphics[width=.8\textwidth]{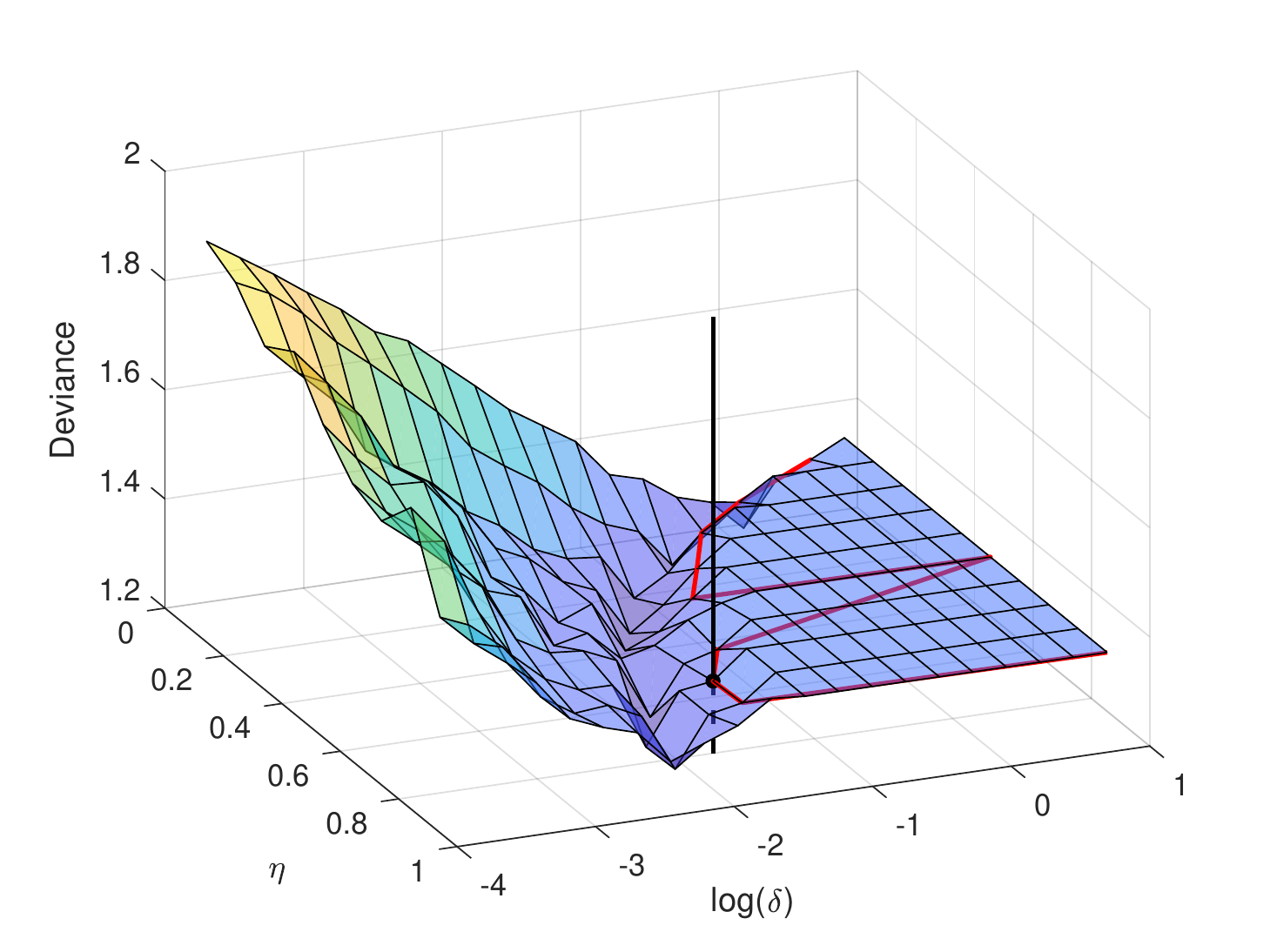}
\caption{Mean deviances of SBLR 5-fold cross validation (CV) on simulated data. The grid of values for $\eta$ is $\{0.1,0.2,\dots, 1\}$ and the red curve indicates the locations of the optimal $\delta$ for each value of $\eta$ such that the mean CV deviance is within one standard-error of the minimum. The black line locates the selected values for $(\delta,\eta)$ corresponding to the smallest deviance along the red curve -- referred to as ``one-standard-error" rule in simulations and applications.}
\label{penalty-CV}
\end{figure}

SBLR model is estimated under $K=5$ and 20 initializations. We use 5-fold cross validation (CV) to tune the penalty factors $\delta$ and $\eta$ in \eqref{elanet-pen} as discussed in Section \ref{other details}. Specifically, the grid of values for $\eta$ is picked as $\{0.1,0.2,\dots, 1\}$. We set $\delta_{\min} = 0.01 \delta_{\max}$ and choose a sequence of 20 equally spaced $\delta$ values on the logarithmic scale. The total runtime for 5-fold CV on this grid is around 16 minutes. Figure \ref{penalty-CV} displays the mean CV deviance on the left-out data and indicates the selected values for $(\delta,\eta)$ under the ``one-standard-error'' rule. The best local solution found at the chosen $(\delta,\eta)$ does not change when increasing to 50 initializations. 

\begin{figure}[htb]
\centering
\includegraphics[width=.32\textwidth]{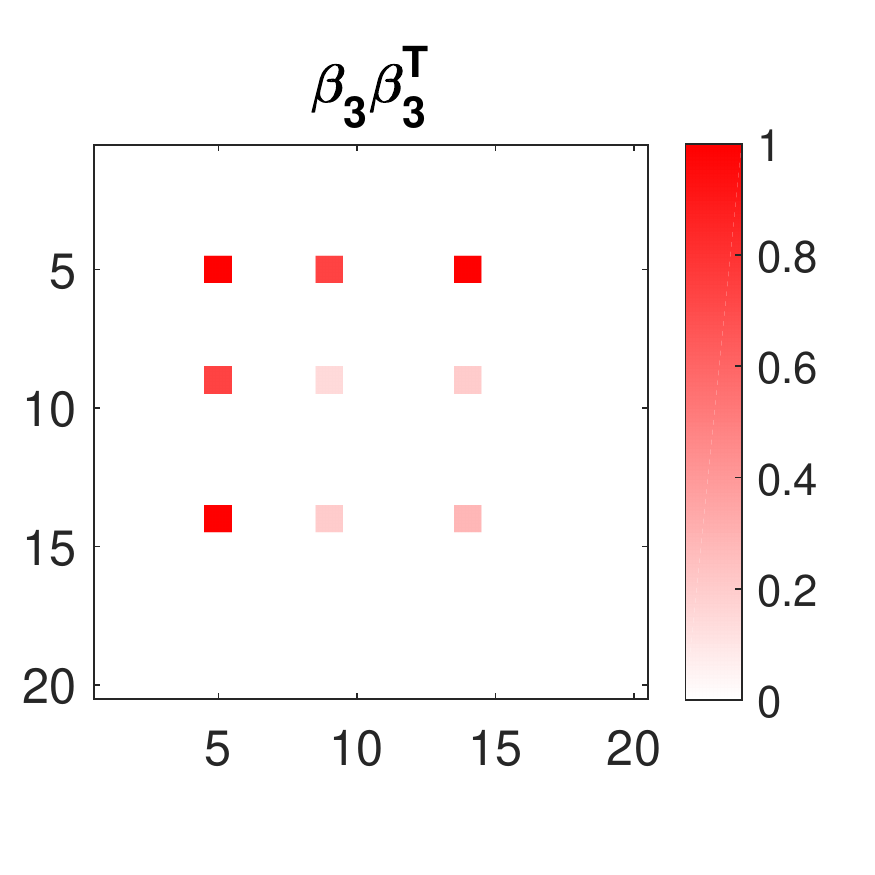}\hspace{5pt} \includegraphics[width=.29\textwidth]{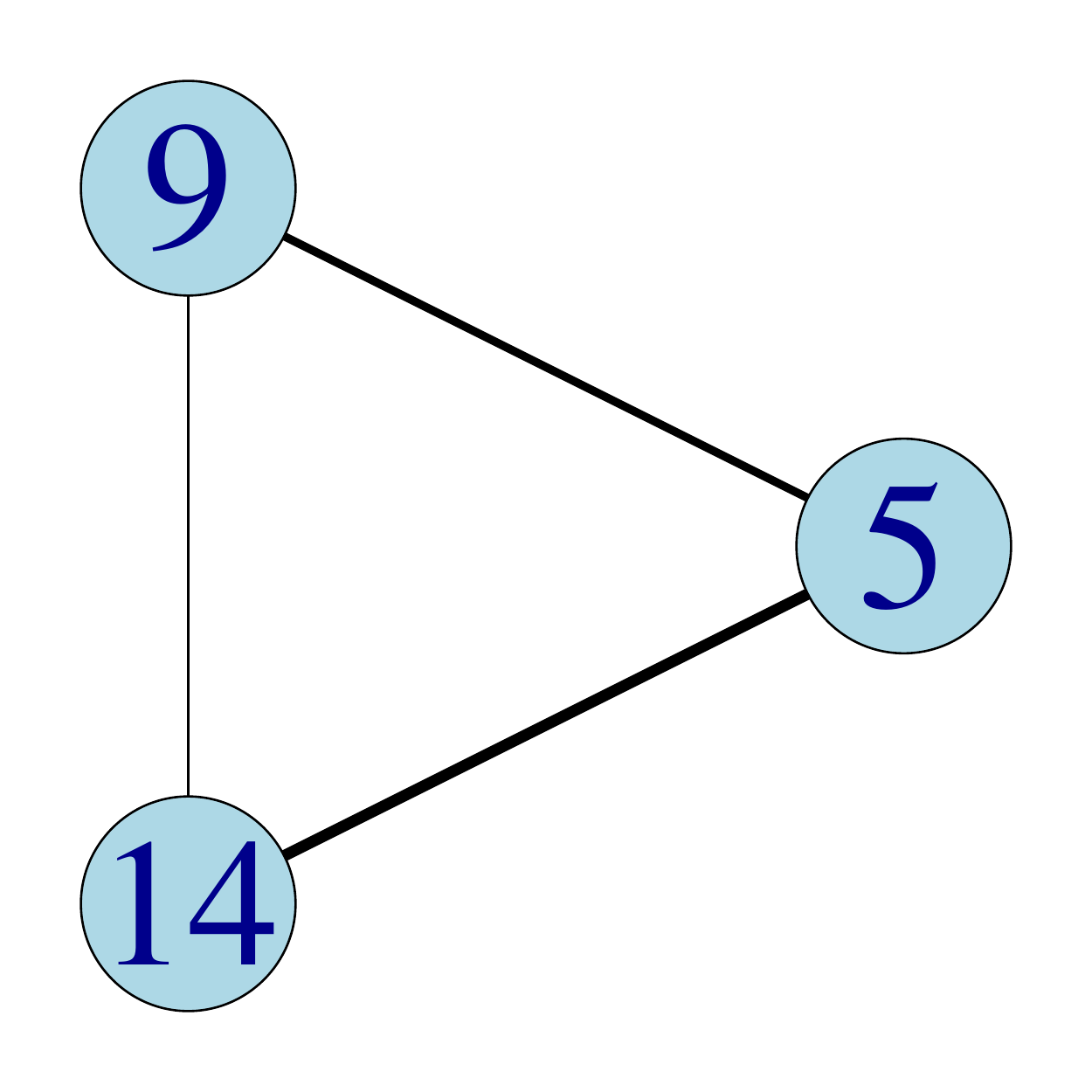} \hspace{5pt} \includegraphics[width=.31\textwidth]{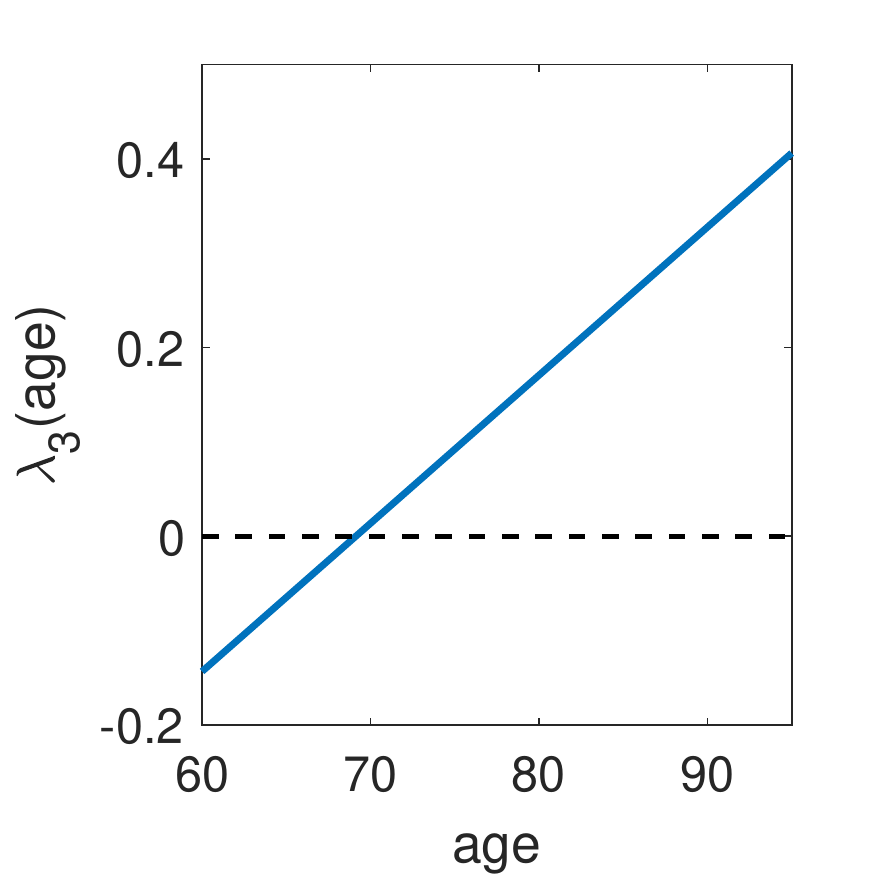}
\caption{Estimated results of SBLR under $K=5$ where the penalty factors are chosen by the one-standard-error rule. Left: the estimated nonzero component matrix $\boldsymbol{\beta}_{3}\boldsymbol{\beta}_{3}^{\top}$; the matrix is normalized such that the off-diagonal element with the largest magnitude is 1. Middle: the corresponding selected subgraph, where the thickness of edges is proportional to the magnitude of their estimated coefficients in $\boldsymbol{\beta}_{3}\boldsymbol{\beta}_{3}^{\top}$; black edges denote true signal edges while red ones falsely identified edges (not displayed). Right: the corresponding estimated age effect $\lambda_{3}(g)$.}
\label{simu-one-std-err-SBLR}
\end{figure}

\begin{figure}[htb]
\centering
\includegraphics[width=.9\textwidth]{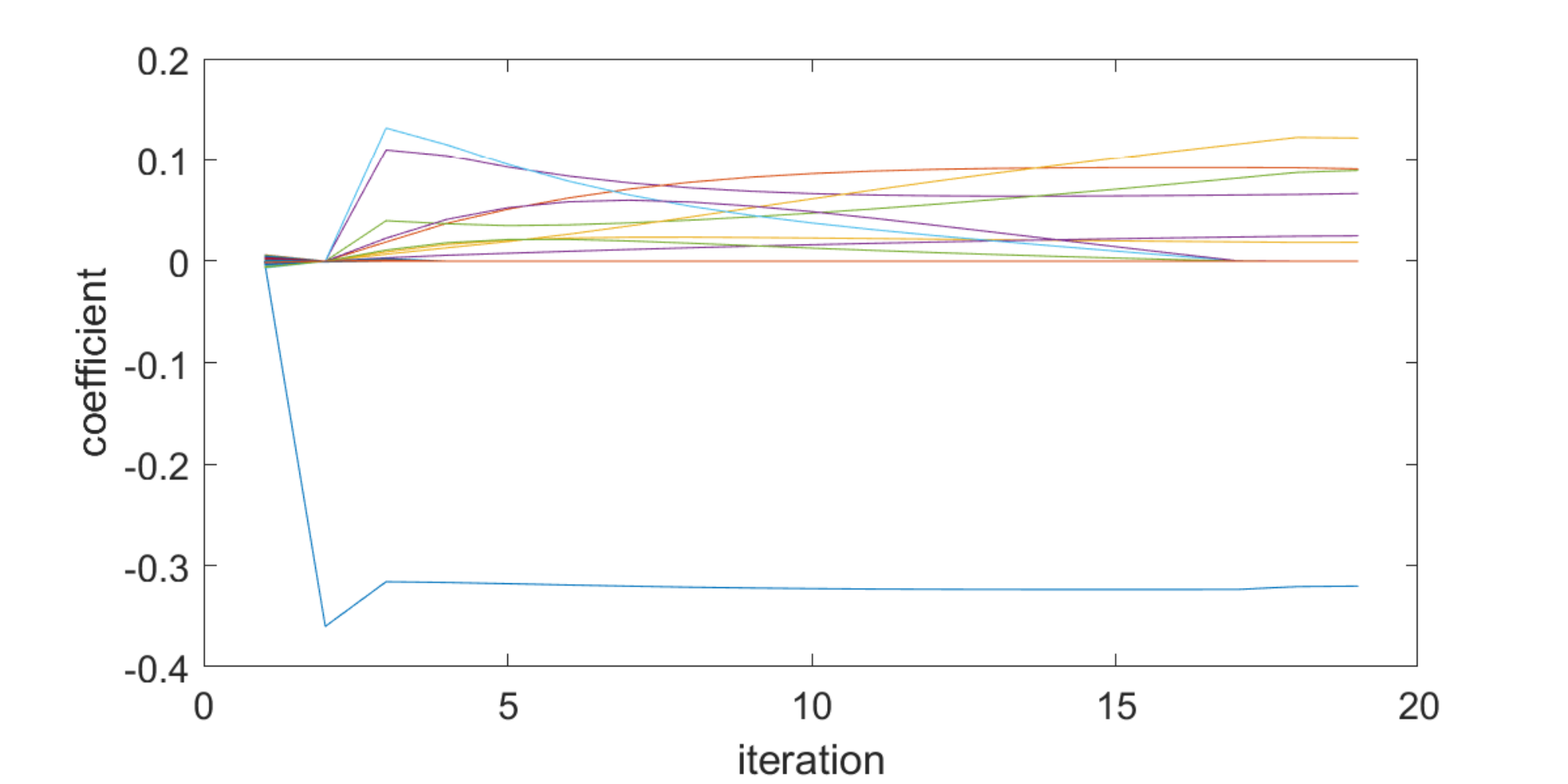} 
\caption{Profiles of estimated entries in coefficient matrices $\left\{ \left(\alpha_{h}\boldsymbol{\beta}_{h}\boldsymbol{\beta}_{h}^{\top},\rho_{h}\boldsymbol{\beta}_{h}\boldsymbol{\beta}_{h}^{\top},\gamma_{h}\boldsymbol{\beta}_{h}\boldsymbol{\beta}_{h}^{\top}\right):h=1,\dots,5\right\} $ over iterations from SBLR at the chosen optimal penalty factors.
}
\label{profile-evolve}
\end{figure}

Figure \ref{simu-one-std-err-SBLR} displays the estimated results of SBLR at the selected values for  ($\delta$, $\eta$) in Figure \ref{penalty-CV}. In this case, SBLR identifies one nonempty component, which partially recovers the first true signal subgraph in Figure \ref{true-signal-subgraphs} with no wrong edges. SBLR also correctly estimates that the effect of this subgraph on the outcome is increasing with age, though starts from a negative value. The profiles of the estimated entries in coefficient matrices $\left\{ \left(\alpha_{h}\boldsymbol{\beta}_{h}\boldsymbol{\beta}_{h}^{\top},\rho_{h}\boldsymbol{\beta}_{h}\boldsymbol{\beta}_{h}^{\top},\gamma_{h}\boldsymbol{\beta}_{h}\boldsymbol{\beta}_{h}^{\top}\right):h=1,\dots,5\right\} $ over iterations from SBLR are shown in Figure \ref{profile-evolve}, which indicates the convergence of the sequences of component coefficients as the loss function \eqref{loss-func} converges.

\begin{figure}[!tb]
\centering
\includegraphics[width=.3\textwidth]{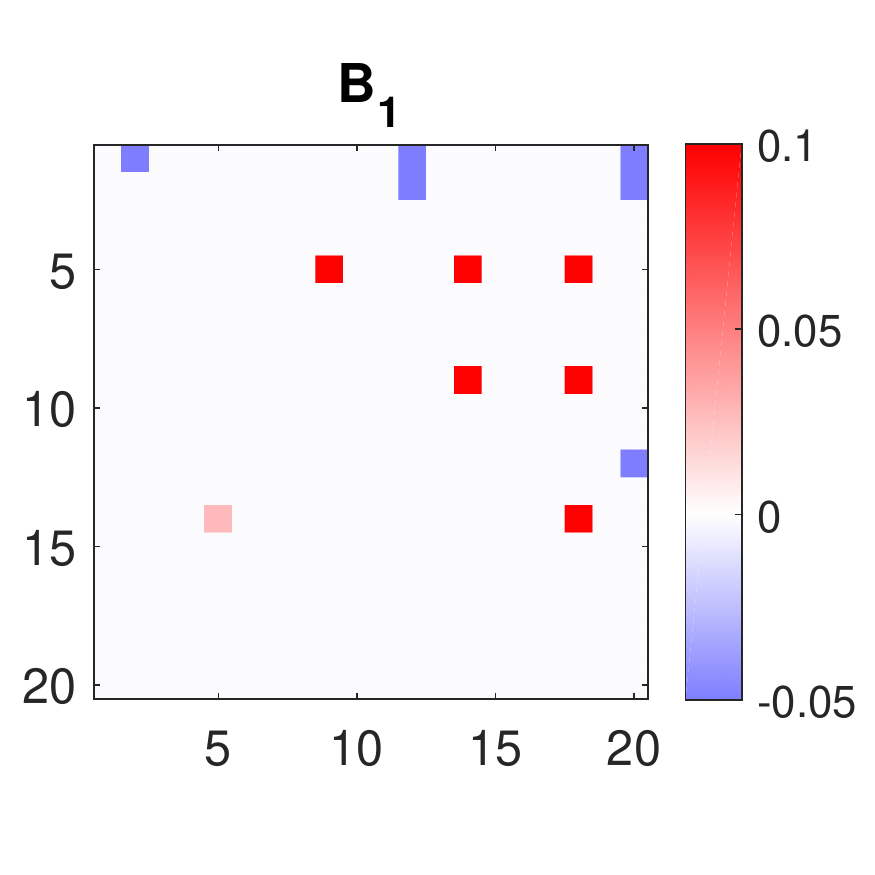}\hspace{5pt} \includegraphics[width=.3\textwidth]{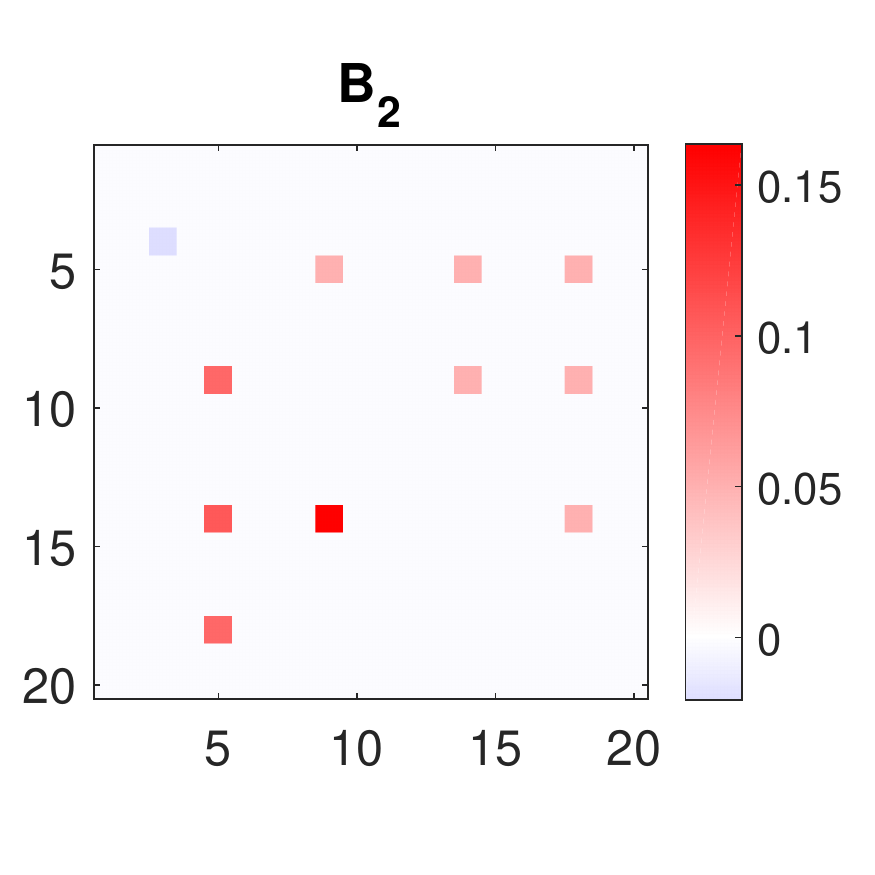}\hspace{5pt} \includegraphics[width=.3\textwidth]{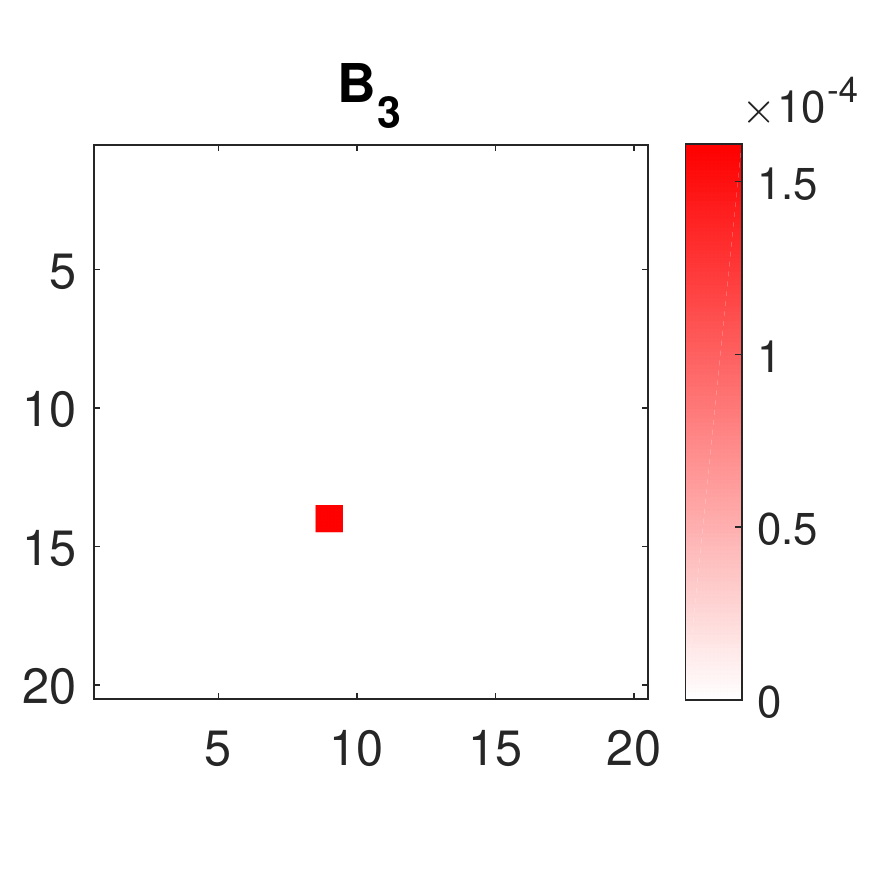} \\
\includegraphics[width=.3\textwidth]{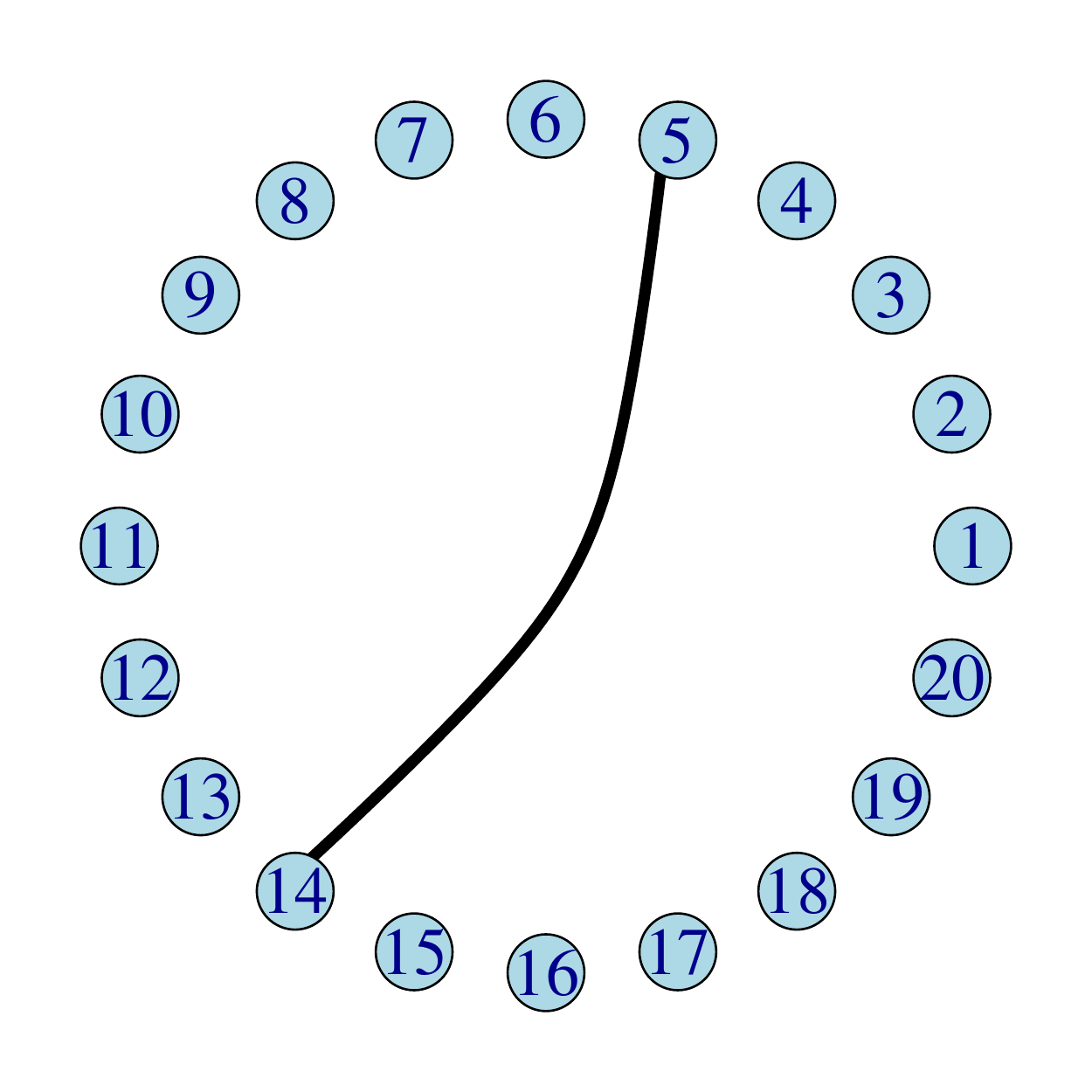}\hspace{5pt} \includegraphics[width=.3\textwidth]{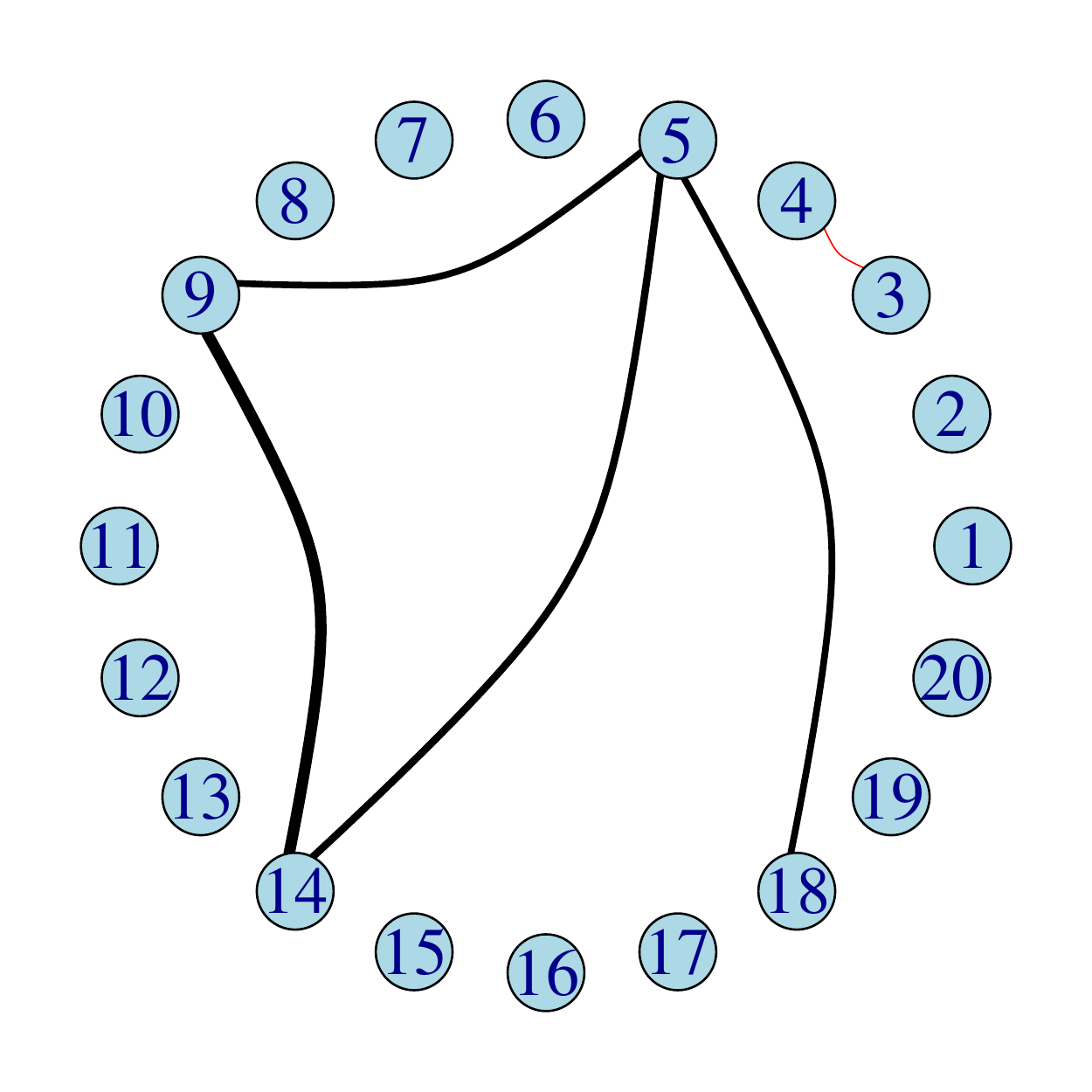}\hspace{5pt} \includegraphics[width=.3\textwidth]{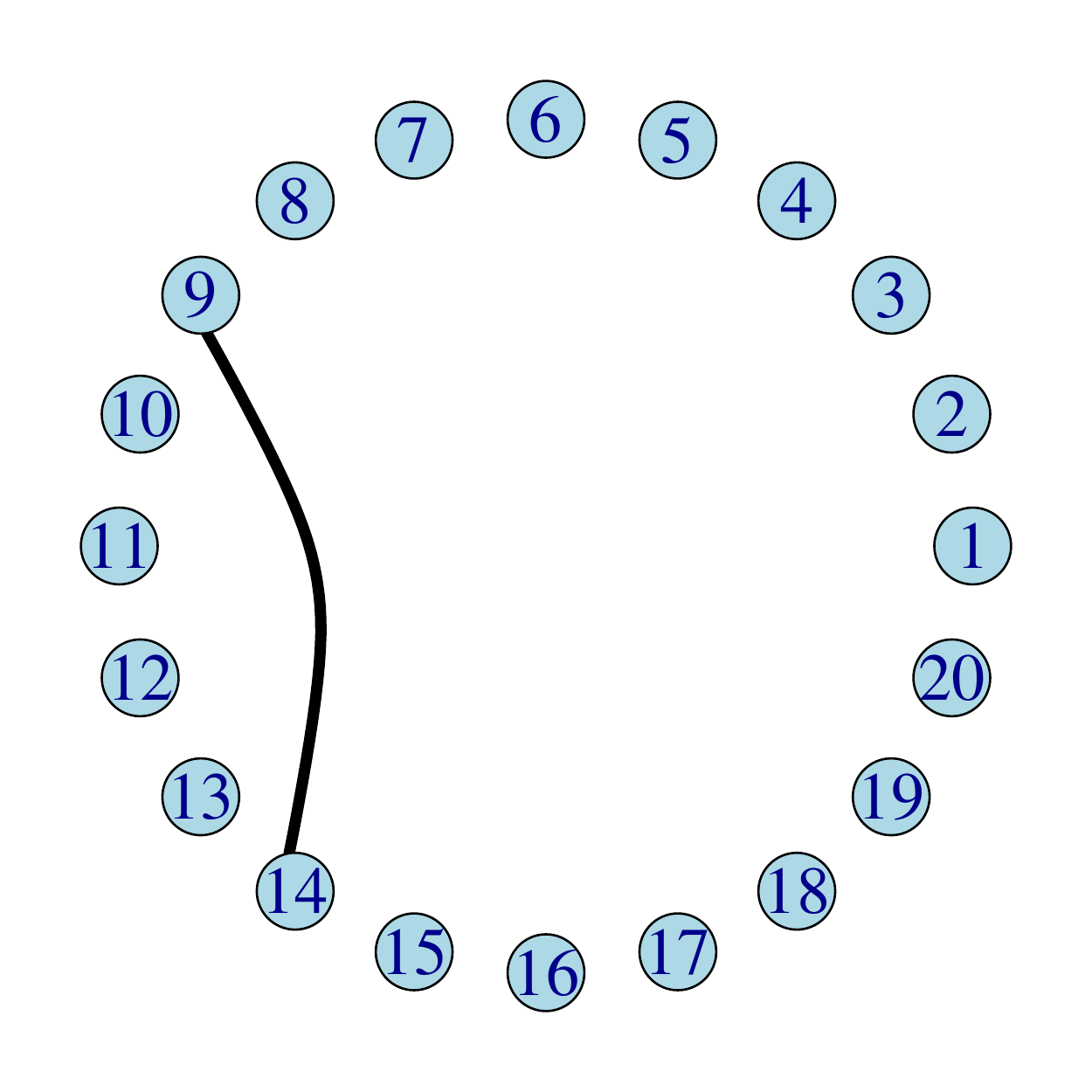} 
\caption{Estimated results of unstructured logistic regression \eqref{unstructure-LR} with elastic-net penalty where the penalty factors are chosen by the ``one-standard-error" rule.  
	Upper panel: estimated entries (lower triangular) versus true values (upper triangular) in $B_1$,  $B_2$ and $B_3$. Lower panel: the selected edges in the network where the thickness of edges is proportional to the magnitude of the corresponding coefficients; black edges denote true signal edges and red ones falsely identified edges.}
\label{simu-one-std-err-LR}
\end{figure}

Figure \ref{simu-one-std-err-LR} shows the estimated results from unstructured LR model \eqref{unstructure-LR}, where the elastic-net penalty factors are also chosen by the ``one-standard-error" rule. Although it is easy to identify a triangle from the selected edges in this simple example, we have to analyze their age effects edge by edge. As shown in Figure \ref{simu-one-std-err-LR}, the selected edge (9, 14) has fake quadratic age effect and there is a falsely identified edge (3, 4).

\begin{figure}[t]
\centering
\includegraphics[width=.33\textwidth]{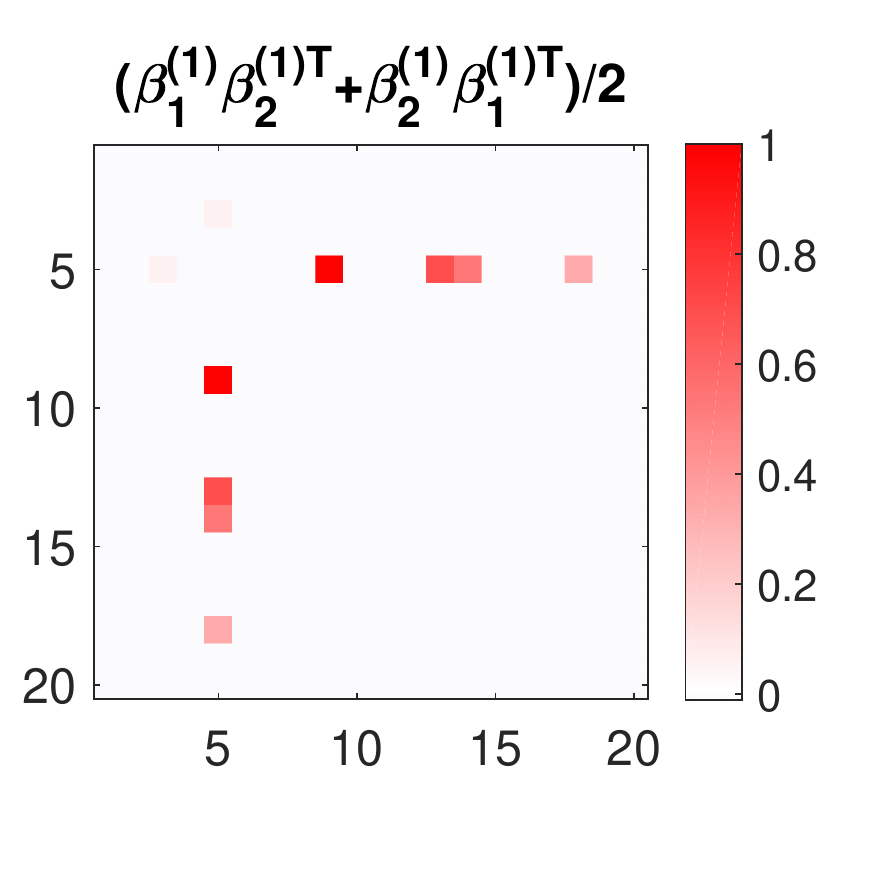}\hspace{5pt} \includegraphics[width=.29\textwidth]{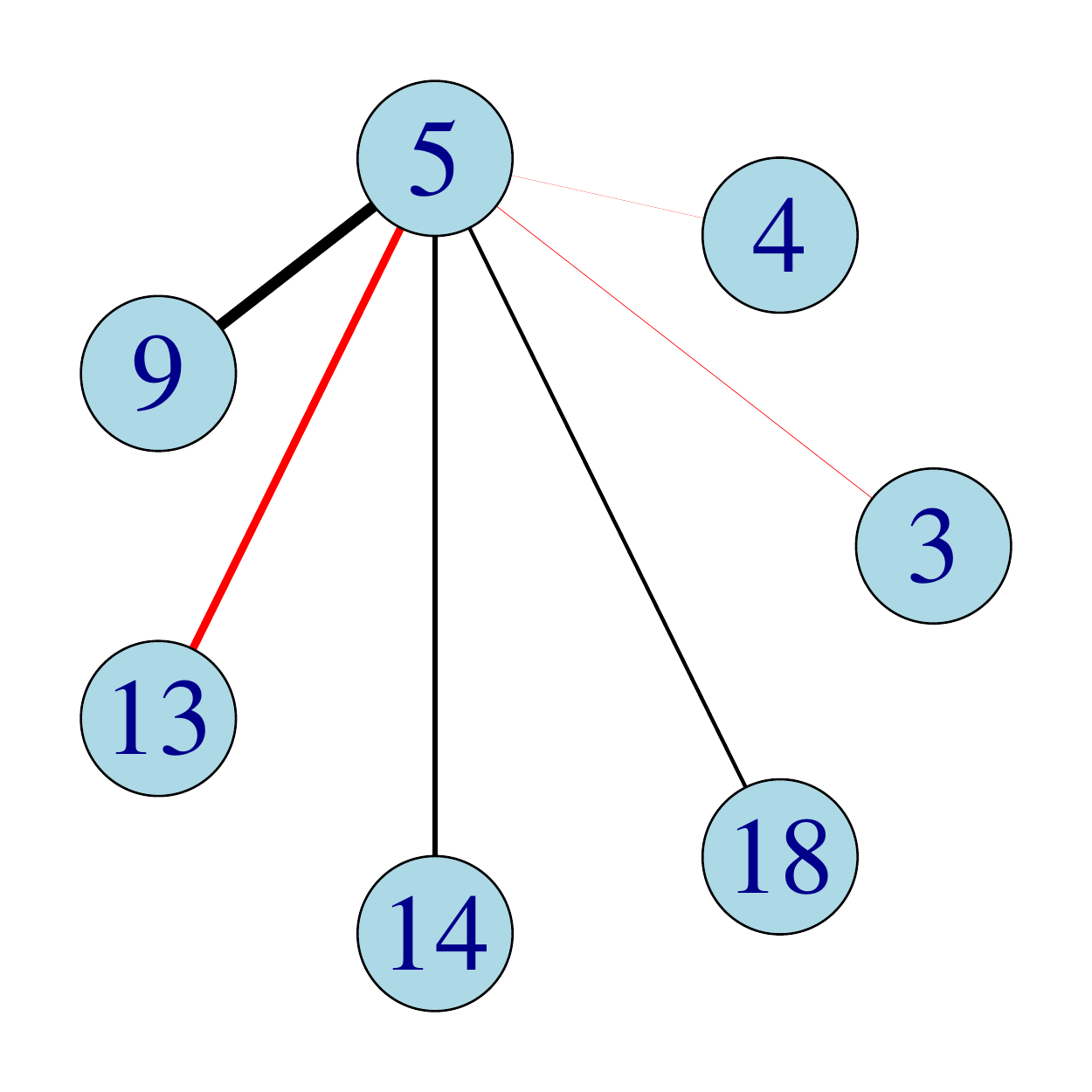} \hspace{5pt} \includegraphics[width=.31\textwidth]{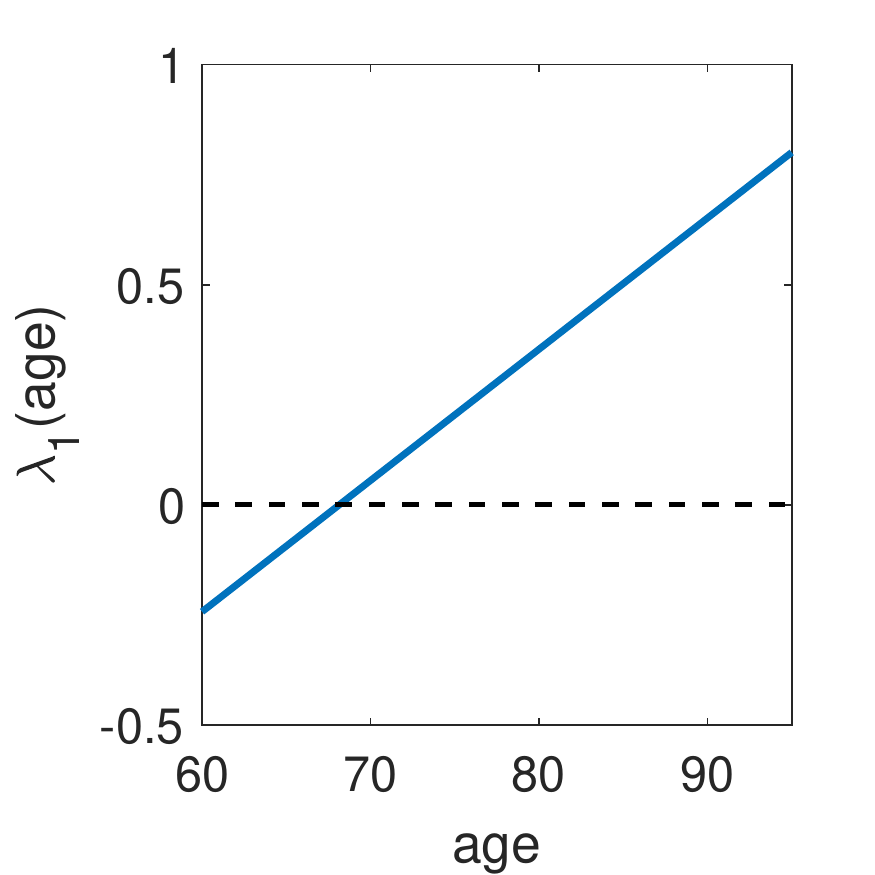}
\caption{Estimated results of NSTR model \eqref{naive-tr} - \eqref{tensor-reg-penalty} under $K=5$ where the penalty factors are chosen by the ``one-standard-error" rule. Left: the estimated nonzero component matrix $(\boldsymbol{\beta}^{(1)}_1\boldsymbol{\beta}^{(1)\top}_2 + \boldsymbol{\beta}^{(1)}_2\boldsymbol{\beta}^{(1)\top}_1)/2$; the matrix is normalized such that the off-diagonal element with the largest magnitude is 1. Middle: the corresponding selected subgraph  where the thickness of edges is proportional to the magnitude of their estimated coefficients in the component matrix; black edges denote true signal edges while red ones falsely identified edges. Right: the corresponding estimated age effect $\lambda_{1}(g)$.}
\label{simu-one-std-err-NTR}
\end{figure}

Figure \ref{simu-one-std-err-NTR} displays the estimated results of NSTR model \eqref{naive-tr} - \eqref{tensor-reg-penalty} under $K=5$ at the optimal penalty factors selected by the ``one-standard-error" rule. As can be seen, NSTR partially identifies the first true signal subgraph in Figure \ref{true-signal-subgraphs} while selecting 3 false edges. The corresponding estimated age effect looks similar to that of SBLR, comparing Figure \ref{simu-one-std-err-NTR} to Figure \ref{simu-one-std-err-SBLR}.

\begin{figure}[!tb]
	\centering
	\includegraphics[width=.32\textwidth]{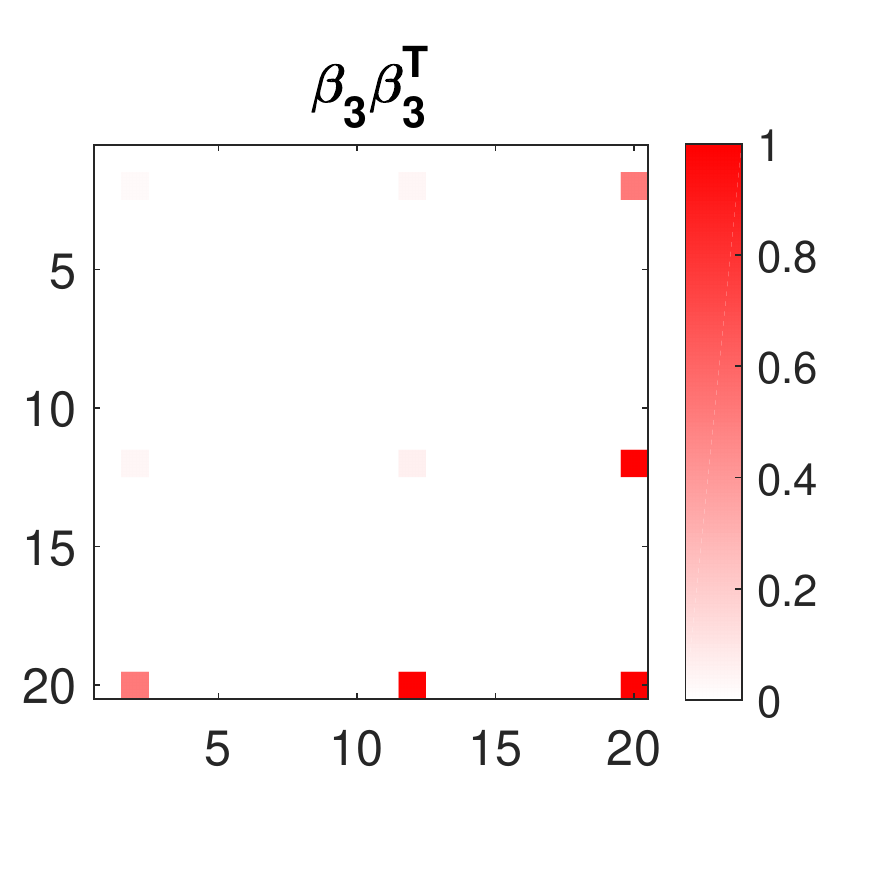}\hspace{5pt}\includegraphics[width=.29\textwidth]{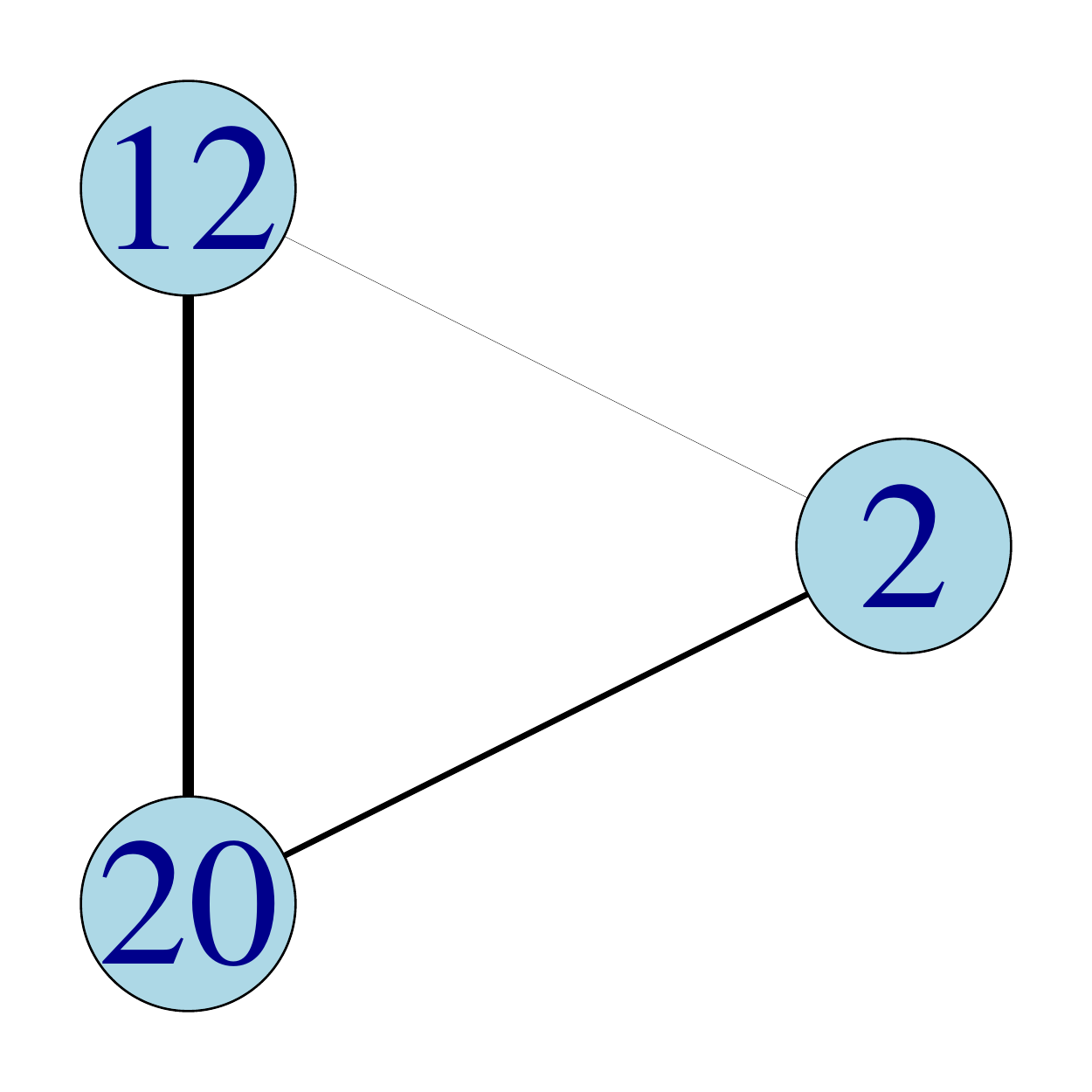}\hspace{5pt}\includegraphics[width=.31\textwidth]{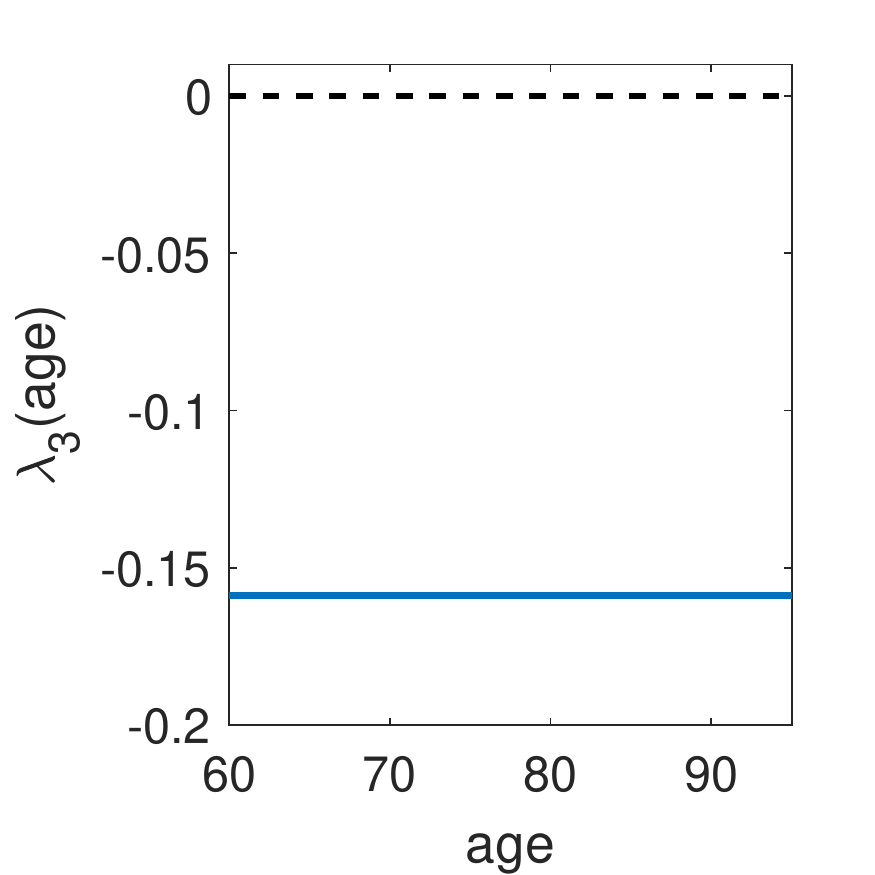} \\ 
	\includegraphics[width=.32\textwidth]{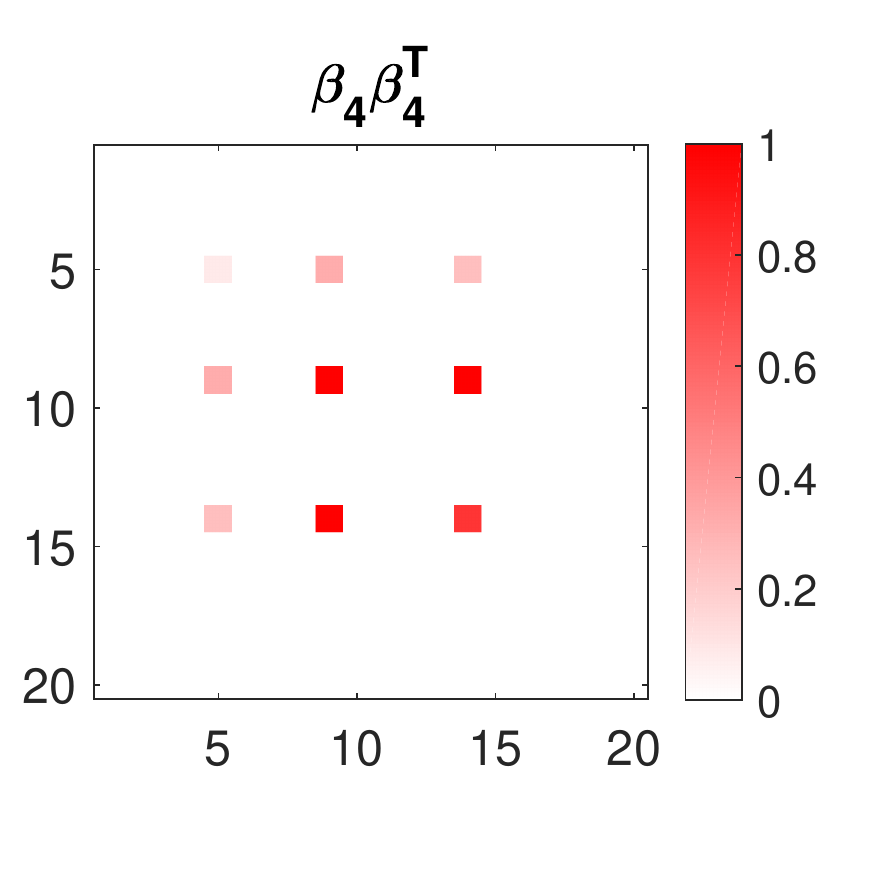}\hspace{5pt}\includegraphics[width=.29\textwidth]{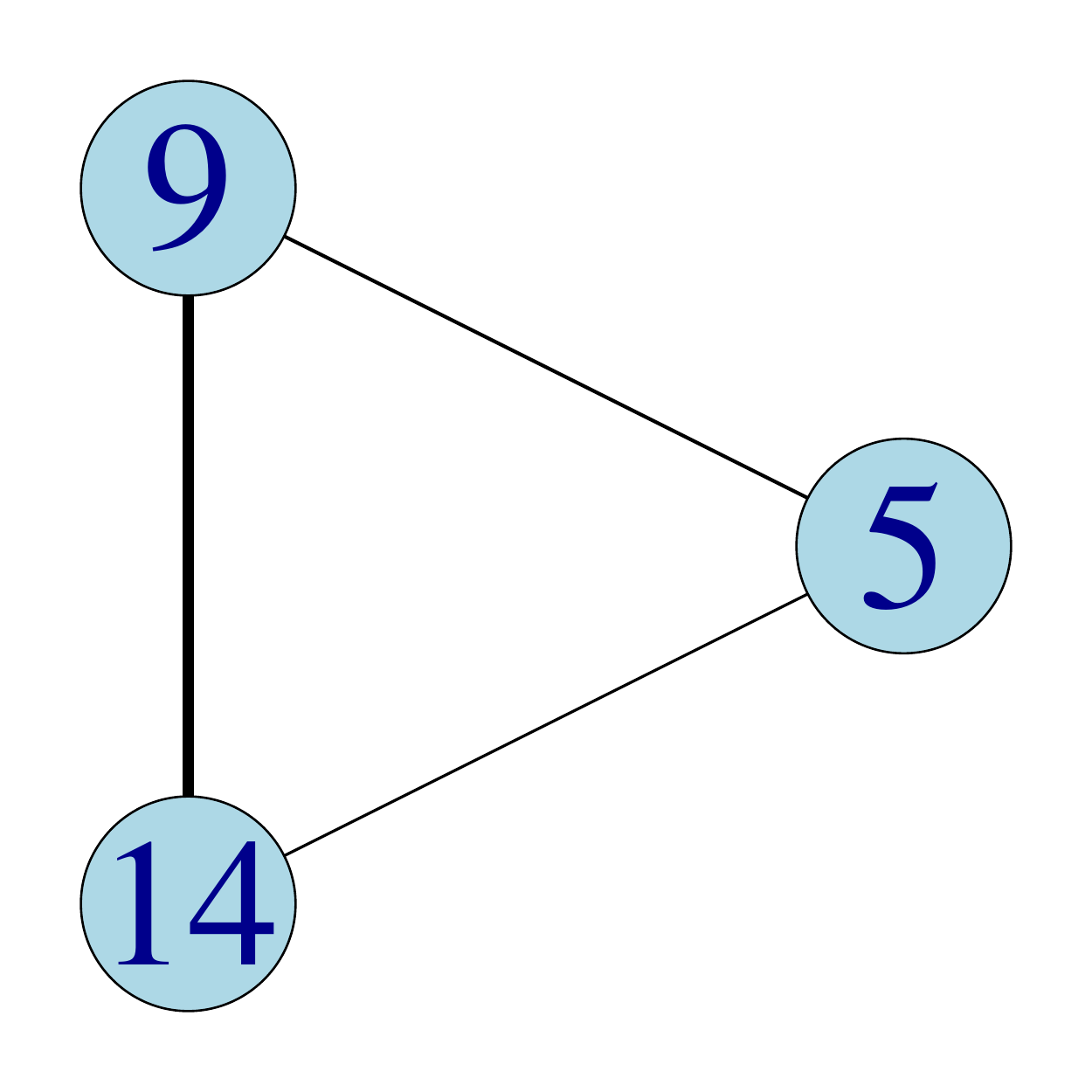}\hspace{5pt}\includegraphics[width=.31\textwidth]{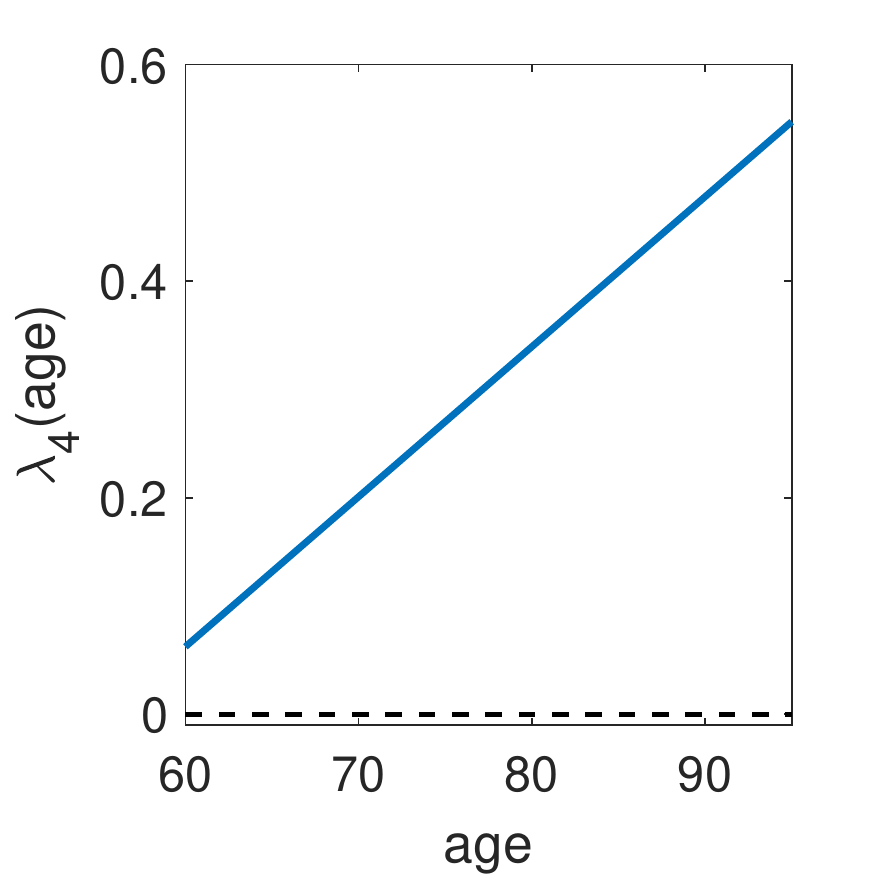} 
	\caption{Estimated results of SBLR on a simulated dataset with $n=1000$ subjects.
		Left: the estimated nonzero component matrices $\boldsymbol{\beta}_{h}\boldsymbol{\beta}_{h}^{\top}$ for $h=3$ (upper) and $h=4$ (lower); the matrices are normalized such that the off-diagonal element with the largest magnitude is 1. Middle: the corresponding selected subgraphs, where the thickness of edges is proportional to the magnitude of their estimated coefficients in the component matrix; black edges denote true signal edges and red ones falsely identified edges (not displayed). Right: the corresponding estimated age effects $\{\lambda_{h}(g)\}$.}
	\label{simu-one-std-err-SBLR1000}
\end{figure}

The performance of SBLR could be improved by increasing the number of observations in the data. Figure \ref{simu-one-std-err-SBLR1000} displays the estimated results of SBLR under $K=5$ when the number of subjects is increased to 1000 under the same data generating process \eqref{generate-networks} - \eqref{generate-response}. This time SBLR partially recovers both true signal subgraphs. The first selected subgraph in Figure \ref{simu-one-std-err-SBLR1000} corresponds to the second true signal subgraph in Figure \ref{true-signal-subgraphs}, where SBLR correctly estimated its effect on the outcome is negatively constant across age. The second selected subgraph in Figure \ref{simu-one-std-err-SBLR1000} is part of the first true signal subgraph in Figure \ref{true-signal-subgraphs}, and SBLR correctly estimated its effect on the outcome is positive and increasing with age.

The procedure described above is repeated 30 times, where each time we generate a synthetic dataset based on \eqref{generate-networks} - \eqref{generate-response}, and record the mean CV deviance at the optimal penalty factors under the ``one-standard-error" rule for SBLR, LR and NSTR. We also record the true positive rate (TPR) for each method, representing the proportion of true signal edges that are correctly identified, and the false positive rate (FPR), representing the proportion of non-signal edges that are falsely identified. Table \ref{all-in-one table} displays the mean and standard deviation (SD) of the mean CV deviance, TPR and FPR for SBLR, LR and NSTR under different problem sizes.

The upper panel of Table \ref{all-in-one table} corresponding to $n=100$ displays the results where each dataset consists of 100 subjects. As can be seen, the average of the mean CV deviance, TPR and FPR from SBLR under $K=10$ are very similar to those under $K=5$, implying that the performance of SBLR is robust to the chosen upper bound for the number of components. In this case, SBLR models achieve competitive predictive performance with LR and better performance than NSTR in terms of out-of-sample deviance. Although SBLR has smaller average TPR than LR or NSTR, it has the lowest average FPR. 

\begin{table}[!tb]
	\centering
	\caption{Mean and SD of the mean CV deviance, TPR and FPR over 30 simulations.}
	\label{all-in-one table}
	\begin{tabular}{cccc}
		\toprule 
		& CV Deviance & TPR & FPR \tabularnewline
		\midrule 
		\multicolumn{4}{c}{$n=100$}\tabularnewline
		\midrule 
		LR & 1.3018$\pm$0.0785 & \textbf{0.3694$\pm$0.2725} & 0.0279$\pm$0.0526\tabularnewline
		\midrule 
		NSTR ($K=5$) & 1.3200$\pm$0.0930 & 0.2316$\pm$0.2718 & 0.0251$\pm$0.0424\tabularnewline
		\midrule 
		SBLR ($K=5$) & \textbf{1.3010$\pm$0.0978} & 0.2189$\pm$0.2705 & 0.0187$\pm$0.0463\tabularnewline
		\midrule 
		SBLR ($K=10$) & 1.3030$\pm$0.0922 & 0.2129$\pm$0.2799 & \textbf{0.0170$\pm$0.0315}\tabularnewline
		\midrule 
		\multicolumn{4}{c}{$n=1000$}\tabularnewline
		\midrule 
		LR & 1.1808$\pm$0.0374 & \textbf{0.7525$\pm$0.1763} & 0.0185$\pm$0.0256\tabularnewline
		\midrule 
		NSTR ($K=5$) & 1.1849$\pm$0.0375 & 0.5710$\pm$0.1894 & 0.0425$\pm$0.0506\tabularnewline
		\midrule 
		SBLR ($K=5$) & \textbf{0.1767$\pm$0.0393} & 0.6374$\pm$0.2437 & \textbf{0.0093$\pm$0.0134}\tabularnewline
		\bottomrule
	\end{tabular}
\end{table}

The lower panel of Table \ref{all-in-one table} shows that increasing the number of observations could improve the performance of SBLR and LR, though may not be the case for NSTR since its mean FPR increases considerably with $n$. When we increase the number of subjects in each dataset from $n=100$ to $n=1000$, the average CV deviance and FPR decrease significantly for SBLR and LR, while their average TPRs increase significantly; the standard deviation of each measure also decreases for SBLR and LR. In the case of $n=1000$, Table \ref{all-in-one table} shows that both average FPR and mean CV deviance of SBLR are smaller than those of LR. Therefore we consider SBLR as a conservative method that tends to be more sparse in selecting signal subgraphs than these competitors.

\section{Application}

Cognitive aging is typically characterized by declines with age in processing speed and memory domains \citep{park2009adaptive}. However some older adults  do not exhibit the expected reduction in cognition, but have superior cognitive performance compared to age- and education-matched cognitively normal older adults \citep{lin2017cingulate}, or even when compared to younger or middle-aged adults \citep{sun2016youthful}. These individuals exhibiting successful cognitive aging are thus called ``supernormals" \citep{mapstone2017success} or ``superagers" \citep{rogalski2013youthful}.


We applied our method to a subset of data from the Alzheimer's Disease Neuroimaging Initiative (ADNI) database as introduced in Section \ref{sec:intro}  to better understand the neural mechanism underlying successful cognitive aging. The dataset contains dMRI data over a 5-year span for 40 supernormals and 45 cognitively normal controls. A state-of-the-art DTI processing pipeline \citep{zhang2018mapping} was applied to extract structural brain networks of subjects. More specifically, we first used a reproducible probabilistic tractography algorithm
\citep{Girard2014266,maier2016tractography} to generate the whole-brain tractography data for  each dMRI scan  in the dataset. The method borrows anatomical information from high-resolution T1-weighted imaging to reduce bias in reconstruction of tractography.  We then used the popular Desikan-Killiany atlas \citep{Desikan2006968} to define ROIs corresponding to the nodes in the structural connectivity network. The Desikan-Killiany parcellation has 68 cortical surface regions with 34 nodes in each hemisphere. Freesurfer software \citep{dale1999cortical,Fischl2004S69} was used to perform brain registration and parcellation. With the parcellation of an individual brain, we extracted two white matter integrity measures - fractional anisotropy (FA) and mean diffusivity (MD), along each fiber tract, and then use the average value to describe connection strength between two ROIs. Both FA and MD are diffusion-related features that characterize water diffusivity along white matter streamlines, and have been widely used in the literature for white matter structure analysis \citep{kraus2007white,jin20173}.

\subsection{Analysis of FA connectivity matrices}

In this case, the weighted adjacency matrix $W_i^{(s)}$ of each brain network consists of mean FA values along fiber tracts between each pair of brain regions. We first compare the predictive performance between SBLR and LR. The penalty factors of elastic-net regularization for both methods are tuned by 5-fold CV as described in Section \ref{other details}. SBLR model is estimated under $K=5$ and 50 initializations, which is sufficient since only one nonempty component is selected under the one-standard-error rule as shown below, and the estimated results are robust when increasing to 100 initializations. The mean CV deviance at the chosen penalty factors of SBLR is 1.29, which is smaller than that of LR, 1.38, indicating better predictive performance.

\begin{figure}[htb]
	\centering
	\includegraphics[width=.32\textwidth]{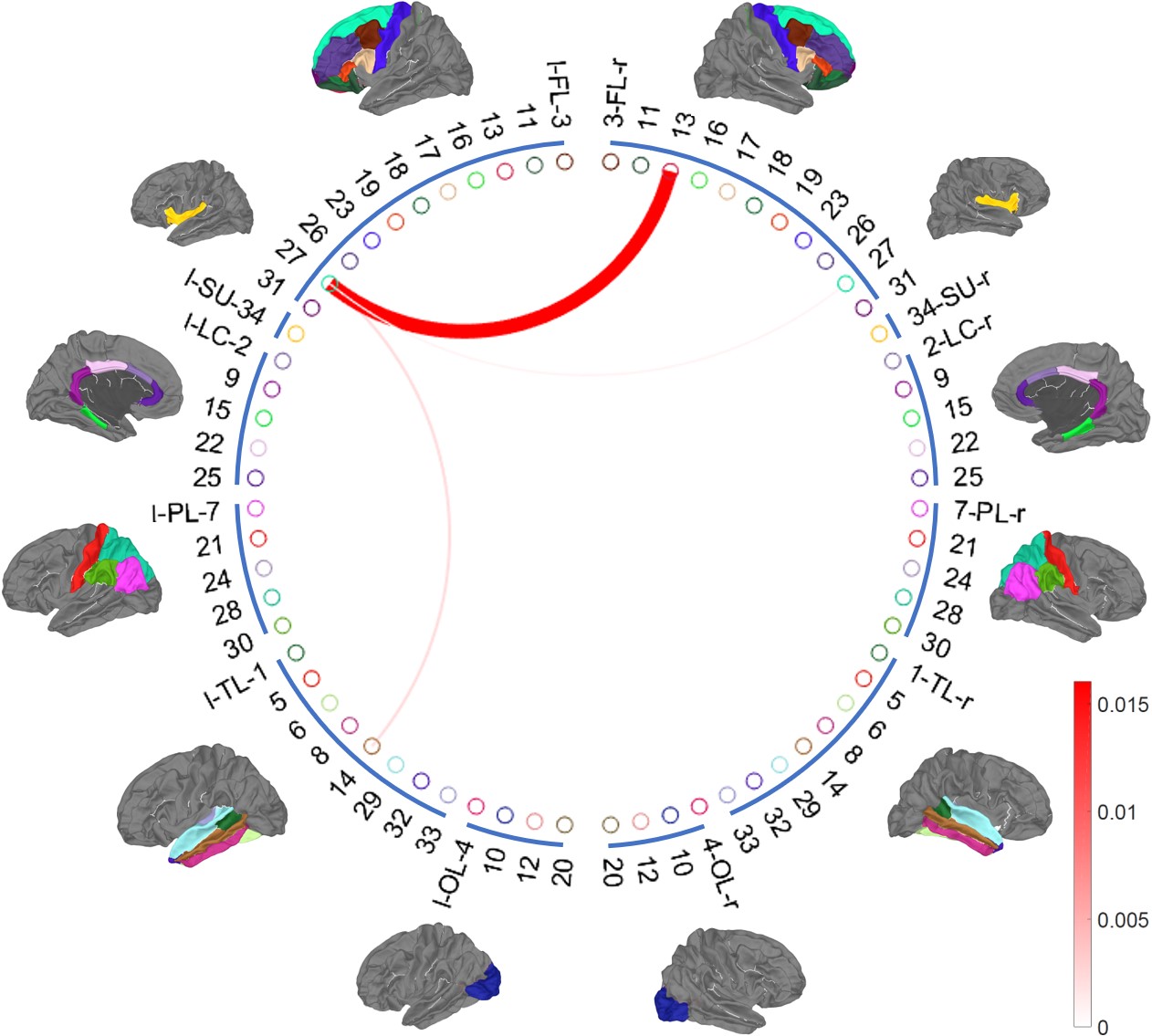} \includegraphics[width=.32\textwidth]{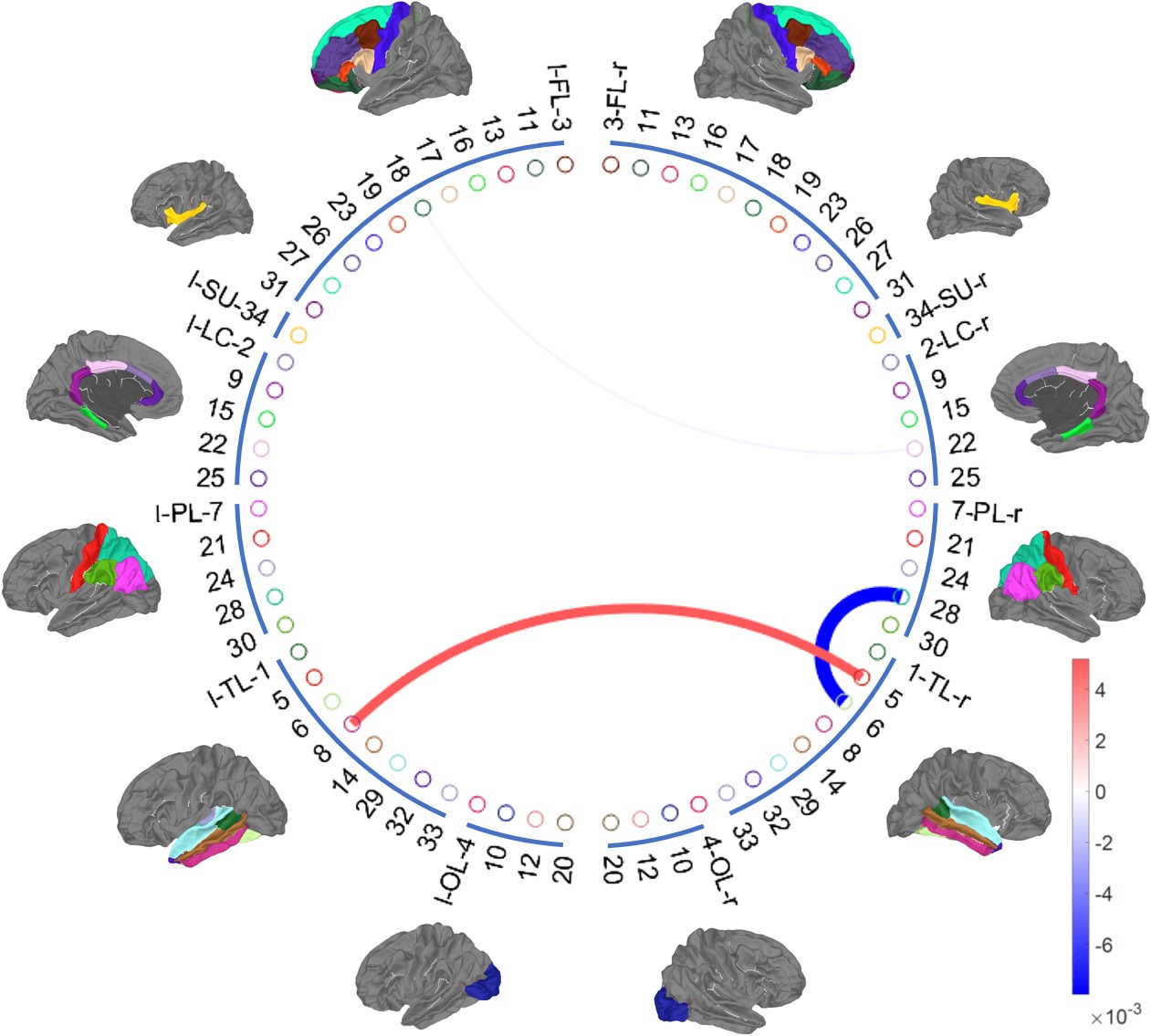} \includegraphics[width=.32\textwidth]{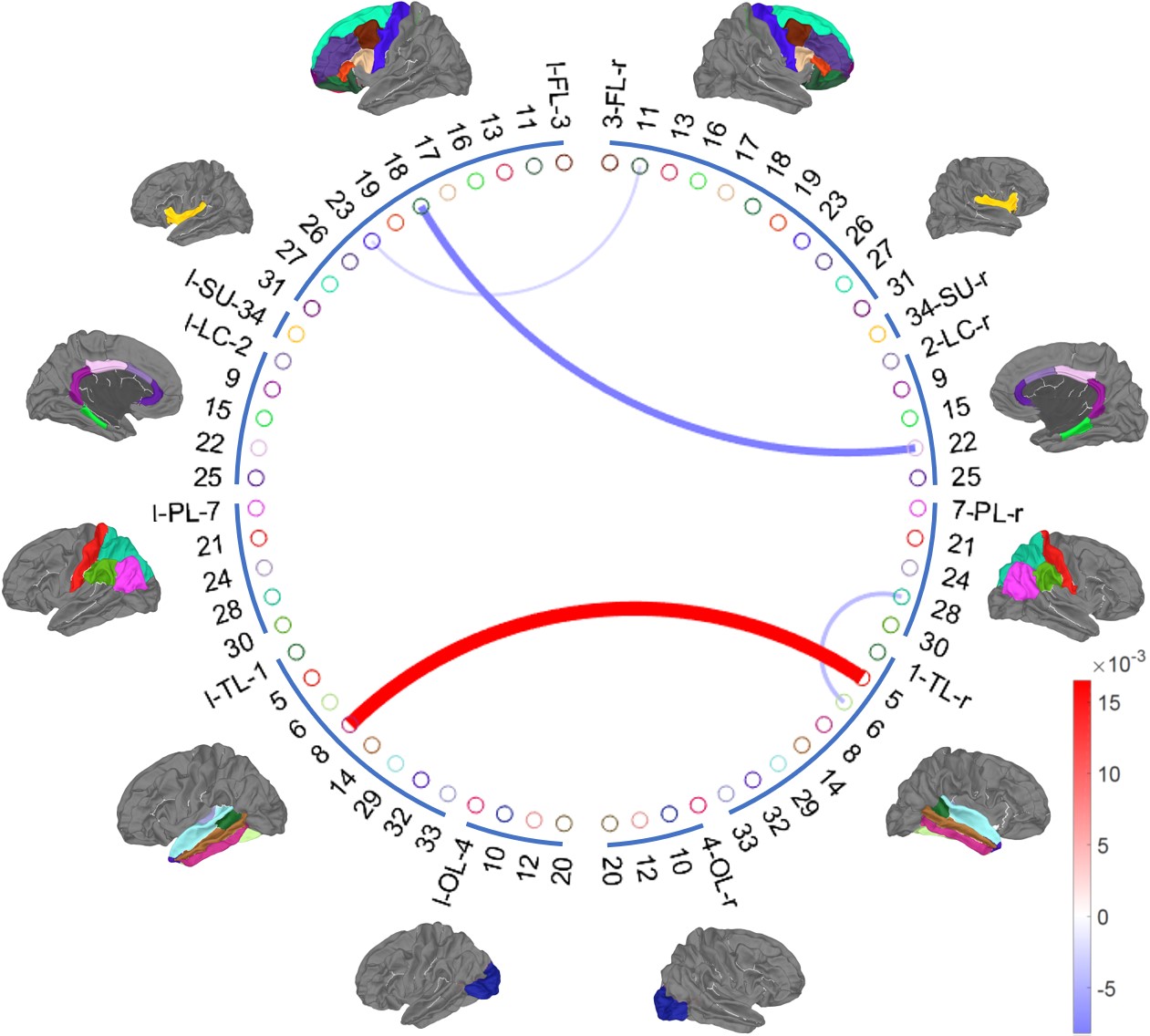}
	\caption{Selected connections predictive of supernormals from LR with constant (left), linear (middle) and quadratic (right) age effect using the FA metric, where the thickness of edges is proportional to the magnitude of the corresponding coefficients, and the color goes from blue to red as the coefficient goes from negative to positive.}
	\label{real-fa-lognet}
\end{figure}

The selected connections predictive of supernormals from LR and the corresponding estimated coefficients in $B_1$, $B_2$ and $B_3$ of \eqref{unstructure-LR} are displayed in Figure \ref{real-fa-lognet}. As can be seen, these connections are scattered in the brain network with no meaningful structure and we have to analyze their age effects edge by edge.

\begin{figure}[htb]
	\centering
	\includegraphics[width=.45\textwidth]{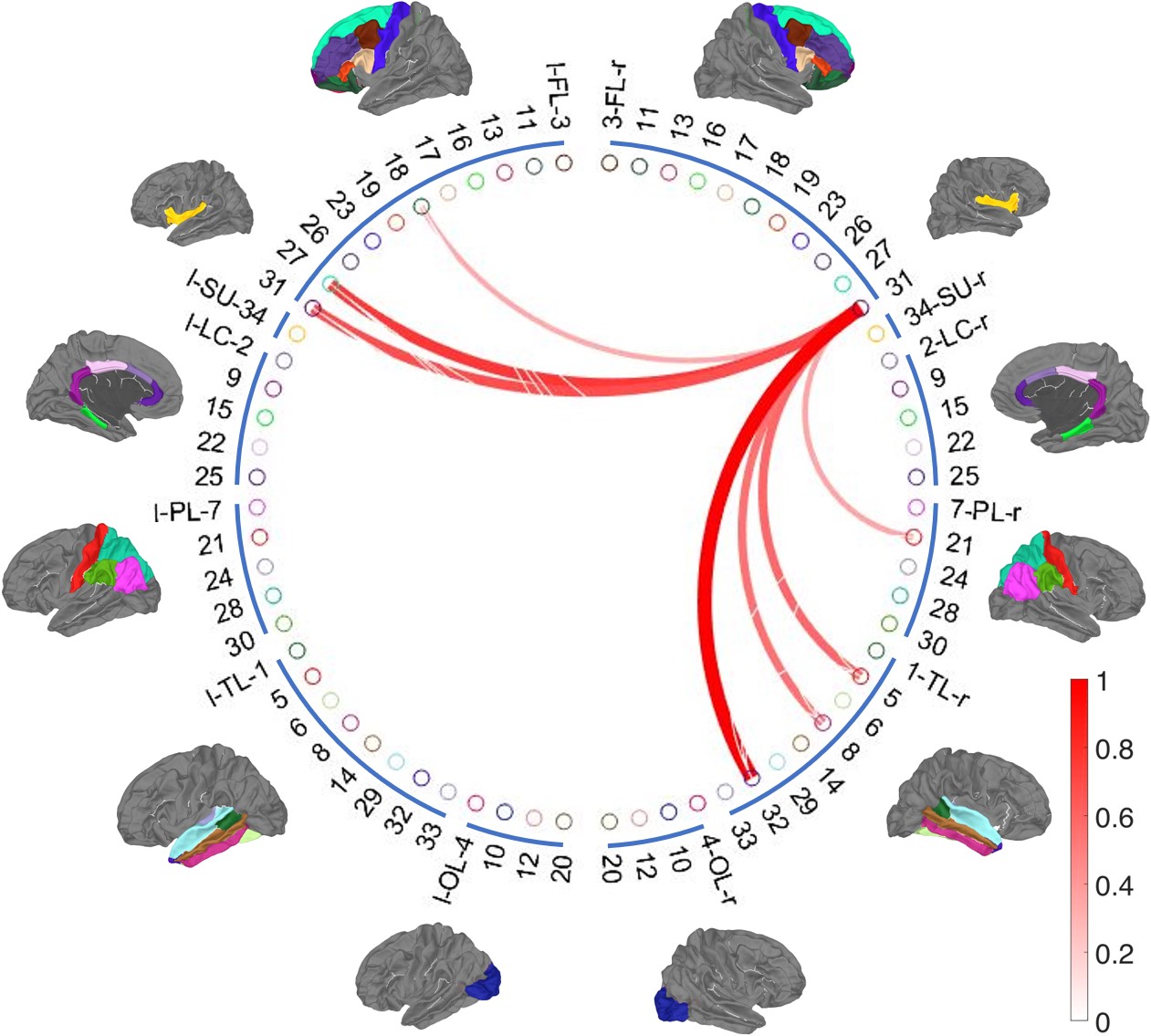} \hspace{5pt} \includegraphics[width=.45\textwidth]{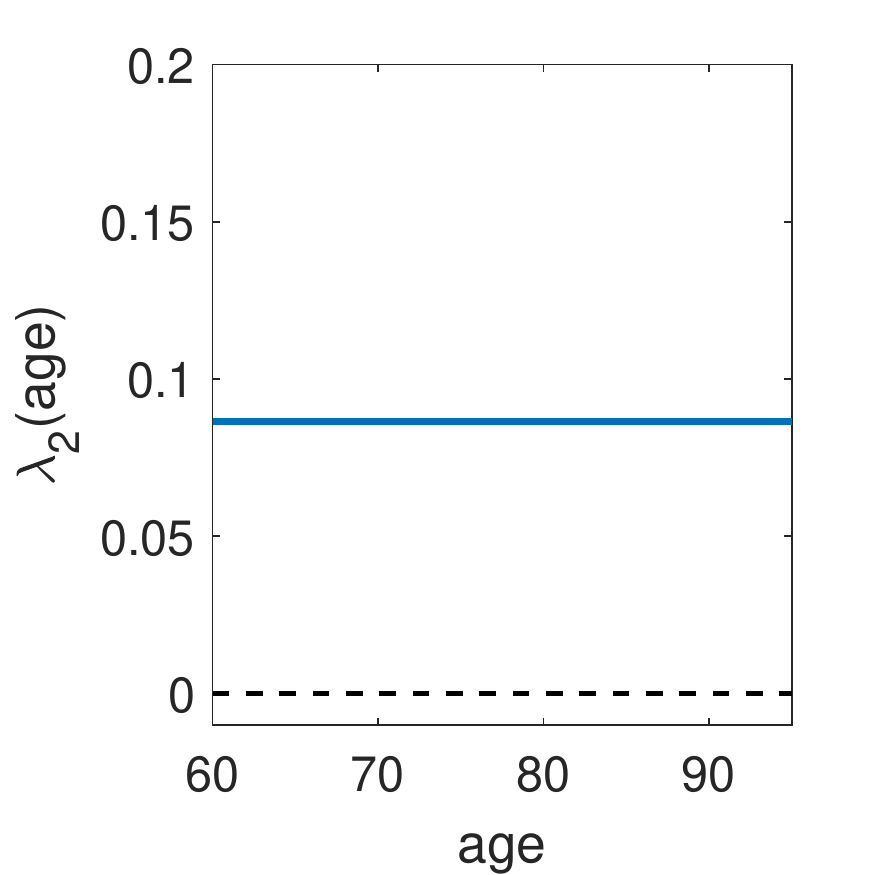}
	\caption{Left: the selected signal subgraph corresponding to the nonzero component $\hat{\boldsymbol{\beta}}_{2}\hat{\boldsymbol{\beta}}_{2}^{\top}$ of SBLR under $K=5$ using FA measures, where the thickness of edges is proportional to the magnitude of their estimated coefficients in $\hat{\boldsymbol{\beta}}_{2}\hat{\boldsymbol{\beta}}_{2}^{\top}$, and the color goes from blue to red as the coefficient goes from negative to positive. Right: the corresponding estimated age effect $\hat{\lambda}_{2}(g)$.}
	\label{real-fa-SBLR}
\end{figure}

SBLR estimated one nonzero component $\hat{\lambda}_{2}(g)\hat{\boldsymbol{\beta}}_{2}\hat{\boldsymbol{\beta}}_{2}^{\top}$
under $K=5$. The nonzero coefficients in $\hat{\boldsymbol{\beta}}_{2}\hat{\boldsymbol{\beta}}_{2}^{\top}$ and the corresponding subgraph are displayed in the left plot of Figure \ref{real-fa-SBLR},  where the matrix $\hat{\boldsymbol{\beta}}_{2}\hat{\boldsymbol{\beta}}_{2}^{\top}$ is normalized such that the off-diagonal element with the largest magnitude is 1.  We notice that the subnetwork SBLR identified consists a hub node, labeled $31r$, in Figure  \ref{real-fa-SBLR}. In the Desikan-Killiany atlas, $31r$ is the frontal pole area, which is also considered as the Brodmann area 10 (BA 10). The other nodes in this subnetwork include $27l$, superior frontal region, and $32r$, temporal pole. Superior frontal region usually controls the sensory system and  temporal pole relates to episodic memories, emotion and socially relevant memory. This subnetwork with BA 10 as a hub is super important, playing an important role in  high level information integration from visual, auditory, and somatic sensory systems to achieve conceptual interpretation of the environment, which is critical to maintain cognition. The right plot of Figure \ref{real-fa-SBLR} shows that the effect of the connection strengths within this subgraph on the outcome is constant across age. The red edges and the positive function $\hat{\lambda}_{2}(g)$ in Figure \ref{real-fa-SBLR} indicate that the effects on the outcome of the connection strengths in the signal subgraph are all positive, implying that older adults with stronger neural connections among these brain regions are more likely to be supernormals.

\subsection{Analysis of MD connectivity matrices}

This time the weighted adjacency matrix $W_i^{(s)}$ of each brain network consists of mean MD values between each pair of brain regions. The mean CV deviance at the chosen penalty factors of LR is 1.39 while that of SBLR is 1.31, again indicating superior predictive performance over the unstructured model.
In this case, LR selected none of the connections predictive of supernormals under the one-standard-error rule.

\begin{figure}[htb]
	\centering
	\includegraphics[width=.45\textwidth]{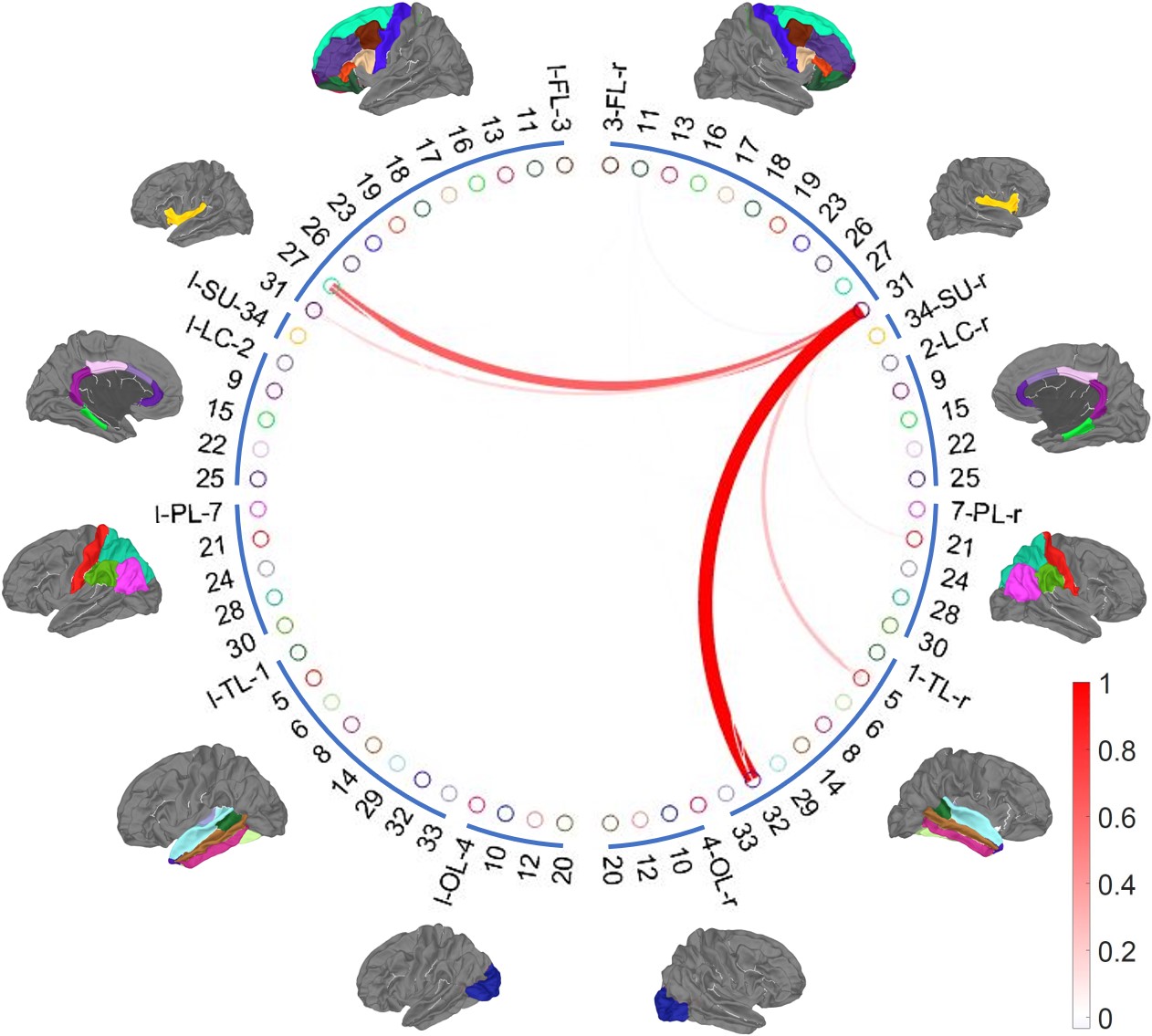} \hspace{5pt} \includegraphics[width=.45\textwidth]{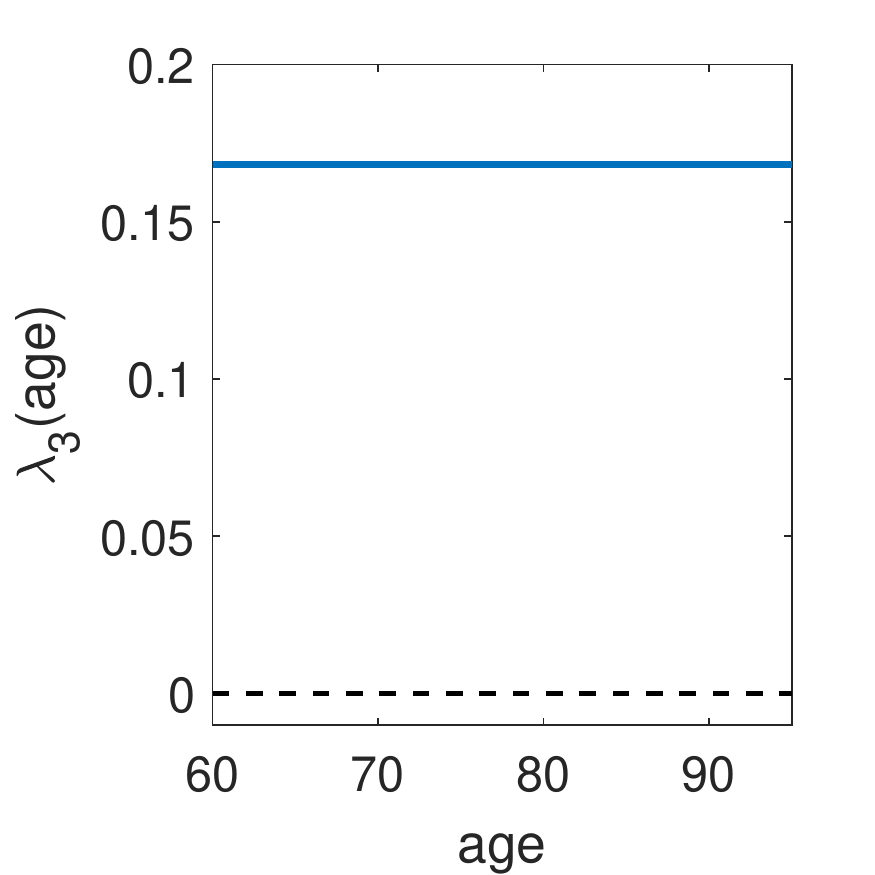}
	\caption{Left: the selected signal subgraph corresponding to the nonzero component $\hat{\boldsymbol{\beta}}_{3}\hat{\boldsymbol{\beta}}_{3}^{\top}$ from SBLR under $K=5$ using MD measures, where the thickness of edges is proportional to the magnitude of their estimated coefficients in $\hat{\boldsymbol{\beta}}_{3}\hat{\boldsymbol{\beta}}_{3}^{\top}$, and the color goes from blue to red as the coefficient goes from negative to positive. Right: the corresponding estimated age effect $\hat{\lambda}_{3}(g)$.}
	\label{real-md-SBLR}
\end{figure}

The signal subgraph identified by SBLR and the corresponding age effect under $K=5$ are displayed in Figure \ref{real-md-SBLR}. As can be seen, the clique signal subgraph selected by SBLR with MD measures is nearly a subgraph of that in Figure \ref{real-fa-SBLR}. The estimated coefficients and age effect also look similar between the two figures. This shows that the estimated results from SBLR are robust to slightly different measures of connection strengths in structural brain networks. 

The involved regions identified in Figure \ref{real-fa-SBLR} and Figure \ref{real-md-SBLR} are also in accordance with the age-related studies in neuroscience. The frontal poles ($31r$, $31l$) and the superior frontal gyrus ($27l$) are among the regions with the greatest age-related reduction in volume and surface area; the right temporal pole ($32r$) has significantly greater-than-average reduction in volume \citep{Lemaitre2012normal}. In addition, lesions in the entorhinal cortex ($5r$) is associated with impairment of episodic memory and the decline in the volume of this cortex predicts progression from mild cognitive impairment to dementia \citep{Rodrigue956}. The constant age effect of the selected subgraph from SBLR under both measures implies that there is no specific age between 60 and 95 that has particularly large predictive effect on the outcome (supernormal or normal aging).

\section{Conclusion}

In summary, the symmetric bilinear logistic regression (SBLR) is a useful tool in analyzing the relationship between a binary outcome and a sequence of longitudinal network observations. SBLR contributes to an insightful understanding of the sub-structure of a network relevant to a binary outcome, as it produces much more interpretable results and lower FPR than unstructured variable selection methods do, while maintaining competitive predictive performance. A coordinate descent algorithm is developed for estimating SBLR with elastic-net penalty, which outputs reliable outcome-relevant subgraphs and their time effects. 
Simulation studies show that the accuracy of SBLR estimates could be improved by increasing the number of subjects in the data. With larger number of longitudinal network observations for each subject (e.g. $\geq 5$), individual variability around the population-level signal subgraphs or age effects could be described in the model and we choose to save it for future work. The method is also straightforward to adapt to regression problems or count responses by a simple modification of the likelihood component in the loss function.


\bibliographystyle{apalike}

\bibliography{SBLR_ref,paperdti}

\end{document}